\documentclass[preprint,onecolumn,amsmath,letterpaper,amssymb,prl,floatfix]{revtex4-1}

\setcitestyle{super}
\usepackage{graphicx}
\usepackage{color}
\usepackage{siunitx}
\usepackage{comment}

\usepackage[sort&compress]{natbib}

\usepackage[scaled]{helvet}
 
\usepackage[T1]{fontenc}

\begin{document}

\title{Iterative Annealing Mechanism Explains the Functions of the GroEL and RNA Chaperones}
\author{D. Thirumalai$^1$, George H. Lorimer$^2$, and Changbong Hyeon$^3$}
\affiliation{$^1$ Department of Chemistry, The University of Texas at Austin, Texas, 78712\\
$^2$ Biophysics Program, Institute For Physical Science and Technology, University of Maryland, College  Park, MD 20742\\
$^3$ Korea Institute for Advanced Study, Seoul 02455, Korea\\
}
\date{\today}
\begin{abstract}
Molecular chaperones are ATP-consuming biological machines, which facilitate the folding of proteins and RNA molecules that are kinetically trapped in misfolded states for long times.  Unassisted folding occurs by the kinetic partitioning mechanism according to which folding to the native state, with low probability as well as misfolding to one of the many metastable states, with high probability, occur rapidly on similar time scales. GroEL is an all-purpose stochastic machine that assists misfolded substrate proteins (SPs) to fold.  The RNA chaperones (CYT-19) help the folding of ribozymes that readily misfold.  GroEL does not interact with the folded proteins but CYT-19 disrupts both the folded and misfolded ribozymes. Despite this major difference, the Iterative Annealing Mechanism (IAM) quantitatively explains all the available experimental data for assisted folding of proteins and ribozymes. Driven by ATP binding and hydrolysis and GroES binding, GroEL undergoes a catalytic cycle during which it samples three allosteric states, referred to as $T$ ({\it apo}), $R$ (ATP bound), and $R^{\prime\prime}$ (ADP bound). In accord with the IAM predictions, analyses of the experimental data shows that the efficiency of the GroEL-GroES machinery and mutants is determined by the resetting rate $k_{R^{\prime\prime} \rightarrow T}$, which is largest for the wild type GroEL. Generalized IAM accurately predicts the folding kinetics of {\it Tetrahymena} ribozyme and its variants. Chaperones maximize the product of the folding rate and the steady state native state fold by driving the substrates out of equilibrium. Neither the absolute yield nor the folding rate is optimized. 
\end{abstract}

\maketitle

\section{Introduction}

Molecular chaperones have evolved to facilitate the folding of proteins that cannot do so spontaneously under  crowded cellular conditions \cite{Lorimer89Nature,Sigler98AnnRevBiochem,Thirumalai01ARBB}. This important task is accomplished without chaperones imparting any additional information beyond what is contained in the amino acid sequence. Furthermore, chaperones assist the folding of proteins whose folded structures bear no relationship to one another. In other words, chaperones are ``blind" to the architecture of the  folded proteins.  Most of the protein chaperones belong to the family of heat shock proteins (HSPs) that are over expressed when the cells are under stress.  Among the many classes of chaperones, the bacterial chaperonin, GroEL, has been most extensively investigated, possibly because it was the first one to be discovered \cite{Ellis87Nature,Lorimer01PlantPhysiol,Hemmingsen88Nature}. 
Although less appreciated, RNA chaperones have also evolved  to enable the folding of ribozymes \cite{Mohr02Cell,Woodson10RNABiol}, which are readily kinetically trapped implying that {\it in vitro} only a very small fraction folds to the functionally competent state on a biologically relevant time scale \cite{Pan1997JMB}.   Both GroEL and RNA chaperones (CYT-19), which we will collectively  refer to as molecular chaperones from now on, are not unlike molecular motors, such as kinesin, myosin, and dynein. There are many similarities between motors and chaperones. (i) Both motors and chaperones are enzymes that undergo a catalytic cycle, which involves binding and hydrolysis of ATP.  Molecular motors hydrolyze one ATP per step, thus converting chemical energy to mechanical work in order to walk on the linear cytoskeletal filaments (actin or microtubule). Both GroEL and  RNA chaperones consume copious amounts of ATP (see below). They couple the hydrolysis of ATP to perform work by partially unfolding the misfolded RNA or proteins. Indeed, helicase activity is attributed to RNA chaperones, such as CYT-19. Helicases are  biological machine that separate double stranded DNA or RNA and translocate on single stranded nucleic acids.  (ii) During the catalytic cycle, the enzymes (motor head in the case of motors and the subunits in the GroEL particle) undergo spectacular conformational changes, which are transmitted allosterically (action at a distance) throughout the complex (see \cite{Thirumalai19ChemRev} for a recent review). Indeed, it is impossible to rationalize the functions of motors or chaperones without allosteric signaling, which we illustrate more fully for GroEL in this article. (iii) Some of the rates in the catalytic cycles of molecular motors also depend on the presence of actin or microtubule. Similarly, ATPase functions of GroEL are stimulated in the presence of substrate proteins (to be referred to as SPs from now on).  For these reasons, a  quantitative understanding of the functions of molecular chaperones mandates that they be treated as molecular machines.
\\
 
\subsection{GroEL-GroES machine} The complete chaperonin system consists of GroEL, the co-chaperonin GroES, which together form both a 2:1 \cite{Xu97Nature} and 1:1 \cite{Fei13PNAS} complex depending on whether SPs are present or absent.  For it to function, which means assist in the folding of a vast number of SPs that otherwise could aggregate, it requires MgATP as well. The availability of a number of structures that GroEL  visits during the catalytic cycle \cite{Xu97Nature,Fei13PNAS,Fei14PNAS} and theoretical developments \cite{Todd96PNAS,Thirumalai01ARBB} have made it possible to obtain insights into the function of GroEL-GroES system. GroEL, assembled from seven identical subunits,  is a homo oligomer with two rings that are stacked back-to-back, which confers it in an unusual rare  seven fold symmetry in the resting ($T$ or taut) state. Major changes in the structures take place between the allosteric ($T$, $R$, and $R^{\prime\prime}$) states in response to ATP and GroES binding (Figs.\ref{transitions} and \ref{structure}a). The dynamics of allosteric changes in GroEL has been reviewed recently \cite{Gruber16ChemRev,Thirumalai18PhilTrans,Gruber18PhilTransRoySocB,Thirumalai19ChemRev}.  The ATP binding sites are localized in the  base of GroEL corresponding to the equatorial (E) domain, connecting the two rings (Fig. \ref{structure}a).   The E carries bulk (roughly two thirds) of the inertial mass of a single subunit.  Binding sites for the co-chaperonin GroES are localized in the apical (A) domain, which also coincides with the region of interaction of the SPs with  GroEL in the $T$ state. 
We present a schematic in Fig. \ref{Landscapes_KPM}b of the reaction cycle in a single ring. 
We ought to emphasize, right at the outset, that recent advances show that when challenged with SPs the functioning state is the symmetric 14-mer GroEL-GroEs complex, resembling a football \cite{Yang13PNAS,Takei12JBC} (see Fig.\ref{structure}b), and not the asymmetric bullet structure as had been thought for a long time.  

The parts list of this complex machine is GroEL, GroES, MgATP,  and the SPs, that require assistance to reach the folded structures. A few words about the SPs are in order. It has been shown long ago that GroEL is a promiscuous machine that interacts with a vast majority of {\it E. Coli.} proteins as long as they are presented in the misfolded states \cite{Viitanen92ProtSci}. This observation and the subsequent demonstration that most of the  SPs used to study GroEL assisted folding are ones not found in {\it E. Coli.} further buttresses this point. For discussion purposes, we distinguish between permissive and non-permissive conditions. Under permissive conditions, folding to the native state occurs readily {\it in vitro} on a biologically relevant time scale ($\tau_B$), which is between 20--40 minutes for {\it E. coli} proteins at 37$^{\circ}$C. 
Under non-permissive conditions, spontaneous folding does not occur {\it in vitro} with sufficient yield of the folded protein  on the time scale, $\tau_B$. The SPs satisfying this criterion are deemed to be stringent substrates for GroEL. Several {\it in vitro} experiments show that  in most cases the SP folding rates in the wild type GroEL are enhanced only modestly, which is fully explained using theoretical studies \cite{Betancourt99JMB,Baumketner03JMB,Cheung06JMB}. 
In fact, the folding rate could even decrease although this has been shown experimentally \cite{Hoffmann10PNAS} using only a mutant form of GroEL, referred to as SR1, from which GroES does not easily dissociate. In contrast, it is the native yield over a period of time that is maximized \cite{Fedorov97JMB,Todd96PNAS} by the GroEL machinery. 

\subsection{RNA Chaperones} 
The tendency of RNA enzymes (ribozymes) to misfold {\it in vitro} is well established \cite{Treiberscience98,Treiber99COSB,Thirumalai96ACR,RussellJMB01,Woodson05COSB}. Several {\it in vitro} experiments have firmly established that self-splicing ribozymes, such as {\it Tetrahymena} ribozyme, fold to functionally competent state  extremely slowly. Only a very small fraction of the initially unfolded ensemble reaches the folded state rapidly ($\sim$ one second) \cite{Pan1997JMB,RussellJMB99}. The reason for RNAs to be kinetically trapped in metastable states is due to the high stability of the base-paired nucleotides. In addition,  RNA molecules  have considerable homopolymers-like characteristics, which do not fully discriminate between a large number of relatively low free energy structures.  These factors render the folding landscapes of RNAs  more rugged than proteins \cite{Thirum05Biochem}. 
We showed that only $\sim$ 54 \% of RNA secondary structures is made of helices \cite{Hyeon2014IJC} with intact Watson-Crick base pairs, which implies that substantial number of nucleotides are engaged in non-canonical base pairs, bulges, and internal multi loops. 
It is estimated that the life time of a helix made of 6 bps, which gives rise to a free energy barrier of $\delta G^{\ddagger} \approx 10-15$ kcal/mol, can be as large as $\sim 10^5$ sec ($\sim$ 1 month). Thus,  
if a helix forms incorrectly during the folding process, spontaneous unfolding of the helix would not occur on a reasonable time scale. Typical time scales for escape from low free energy some kinetically trapped metastable states can easily exceed hundred minutes (see for example Eq. (2) in \cite{Pan1997JMB}).

The arguments given above suggest that the folding landscape of most RNA molecules ought to consist of multiple metastable minima with similar stability that are separated from the native state by large (compared to $k_B T$ where $k_B$ is the Boltzmann constant and $T$ is the temperature, which unfortunately is the same notation used for the $T$ state in GroEL) free energy barriers \cite{Solomatin10Nature}. The structures in the metastable states often share many features that are common with the folded state. Such rugged landscapes govern the functions of many RNA molecules, such as riboswitches that are involved in transcription and translation. These biological processes are associated with switching between at least two alternative structures.  In riboswitches the switch  between the states  is modulated by  metabolites or metal ions. Of relevance here is the \emph{in vitro} folding of \emph{Tetrahymena} ribozyme, which is a self-splicing intron. 
For this enzyme it is found that 
only a fraction ($\Phi=0.08$) of the initial population of unfolded molecules directly folds to the functionally competent native state in about 1s, and the rest  ($1 - \Phi=0.92$) of the molecules are kinetically trapped in competing basins of attraction \cite{Pan1997JMB,ThirumARPC01,Zhuang00Science} for arbitrarily long times. 
For \emph{Tetrahymena} ribozyme to function, it is essential that several key native tertiary contacts form. 
Incorrect formation of these tertiary contacts leads results in functionally incompetent ribozyme. 
For example, without the formation of the pseudo knot (P3 helix)  
the two domains (P5-P4-P6 and P7-P3-P8) cannot be stabilized (see Fig.\ref{RibozymeFig}a). 
Formation of alternative helix (Alt-P3)  and other misfolded structures impair the function of ribozymes. 
An introduction of a single point mutation (U273A) stabilizing the P3 pseudoknot helix was shown to increase $\Phi$ as high  to 0.8 \cite{Pan00JMB}. 

DEAD-box protein CYT-19, which belongs to a general class of RNA chaperones \cite{Mohr02Cell,Lorsch2002Cell,Tijerina2006PNAS,Bhaskaran07Nature,Grohman2007Biochemistry,Woodson10RNABiol,Mallam2011PNAS,Russell2013RNAbiology},  
comprises of a core helix domain and arginine rich C-terminal tail. Cyt-19 recognizes surface-exposed RNA helices (duplexes) and unwinds them, like helicases belonging to the SF2 family (see Fig.\ref{RibozymeFig}b for the yeast analogue of CYT-19 \cite{Mallam2011PNAS}), into single stranded RNA by expending free energy due to ATP hydrolysis.  It is likely that ATP triggered conformational changes promotes local unwinding of RNA helices. 
Because of the helicase activity of CYT-19, the microscopic mechanism does involve  local unfolding of the accessible helices. Thus, both GroEL and CYT-19 perform work on the misfolded structures by forcibly unfolding them, at least partially.  This is another common theme linking the functions of CYT-19 and GroEL.
 \\

Our goals in this article are the following: (1) We present a unified theoretical perspective on the functions of GroEL and RNA chaperones. The essence of the assisted folding mechanism of the SPs is illustrated using the well investigated GroEL-GroES system. Although the two enzymes, exhibiting machine-like activity, are quite different we show that the theory based on the Iterative Annealing Mechanism (IAM) quantitatively explains a vast amount of experimental data in chaperone-assisted folding of proteins as well as ribozymes. A major conclusion of the theory is that these ATP-consuming chaperones are stochastic machines that drive the SPs or ribozymes out of equilibrium. This implies that in the steady state, $P_{N}^{SS}$, (long time limit) the yield of the folded protein does not correspond to that expected at equilibrium $P_N^{EQ}$, which would be $\propto \exp(-\beta \Delta G_{NU})$ where $\beta = 1/k_B T$, and $\Delta G_{NU}$ is the free energy of stability of the native $N$ state with respect to the unfolded state ($U$). In other words, $P_{N}^{SS} \ne P_N^{EQ}$. (2) The differences between GroEL-GroES system and RNA chaperones  naturally arises from the IAM predictions, and highlights the likely inefficiency (large consumption of ATP relative to the production of the folded state) of RNA chaperones. (3) Because the GroEL structures in different nucleotide states are known, we illustrate the conformational changes that occur during the allosteric transitions in the GroEL  in response to ATP and SP binding and link these changes to the folding of the SPs. (4) Finally, we outline recent developments, which provide incontrovertible evidence for the quantitative validity of the IAM, which establishes that GroEL-GroES system is a parallel processing stochastic machine that simultaneously anneals two misfolded molecules by sequestering one each in the two chambers of the symmetric complex. Remarkably, the symmetric complex forms only when the GroEL-GroES system is subject to load, i.e., challenged with SPs that require assistance to reach the native state. 

\section{Iterative Annealing Mechanism (IAM) for GroEL-GroES}
In this section we systematically develop the physical basis for the IAM by dissecting the fate of the SPs in the absence of the chaperonin machinery. 
We begin by considering how SPs, which do not recruit GroEL-GroES, fold spontaneously.
This is followed by a brief description of the dynamics of allosteric transitions that  GroEL  undergoes in response to  ATP and GroES binding and hydrolysis of ATP.  Lastly, the physical picture of the link between allosteric transitions and SP folding is described, which vividly reveals the machine-like characteristics of the GroEL-GroES system. The applications of the theory of the IAM to  experimental data cement the quantitative validity of the active role GroEL plays in assisted folding.

\subsection{{\it E. Coli} does not have enough GroEL to process the entire proteome} 
Over twenty years ago Lorimer \cite{lorimer1996FASEBJ} showed, using data for {\it E. Coli} B/r growing in minimal glucose medium at 37$^{\circ}$C with a cell doubling time  ($\tau_D$ of $\sim$ 40 minutes, that the number of GroEL particles can only process between (5 -10)\% of the proteome. The crux of the that argument can be summarized as follows. The rate of protein synthesis is $k_S = N_P/\tau_D$ where $N_P$ is the number of polypeptide chains in a cell. Assuming that the total mass of proteins per cell is $\approx$ 1.56$\times$10$^{-13}$ g and the average mass is $\approx 4\times$10$^4$ g/mol then $N_P \approx$ 2.35$\times$10$^6$, which implies that $k_S=6\times10^4$ chains/min. Given there are about 3.5$\times$10$^4$ ribosomes, it follows that this strain of {\it E. Coli} synthesizes about one polypeptide chain every 35 seconds. Needless to say, most if not all of the proteins have to reach the the folded state in the crowded environment without the assistance of GroEL. The average cell contains about $N_{\text{GroEL}}=$1,580 GroEL$_{14}$ particles, and about nearly twice as many GroES molecules. The typical measured values of the rates of assisted folding {\it in vitro}, $k_F$s are in range $(1 - 2)$ min $^{-1}$. Thus, the available GroEL particles can assist in the folding of $N_{\text{GroEL}}k_F \approx$ 3160 polypeptide chains/min, which is clearly far less than the synthesis rate of $6 \times$10$^4$ chains/min. Thus, only about $(3 - 5)$ \% of the proteome can recruit GroEL-GroES in order to fold. Nevertheless, removal of the GroE gene is lethal to the organism, attesting to its importance in {\it E. Coli} growth. These estimates raise the following two important questions: (a) What are the potential SPs that fold with the assistance of the GroEL-GroES system? (b) How do the vast majority of proteins ($\approx$ 95\%) fold without the chaperonin machinery? We will answer the second question here and refer the readers to relevant papers \cite{Stan05ProtSci,Chaudhuri05CellStress,Stan06PNAS,Noivirt-Brik07Bioinformatics,Endo07BiosciBiotechnolBiochem}, except to note that GroEL does not discriminate between proteins based on their folded structures because the very SP residues that interact with GroEL are buried in the folded state \cite{Stan06PNAS}.  

\subsection{Stringent SPs and ribozymes fold by the Kinetic Partitioning Mechanism (KPM)} 
Spontaneous folding of small proteins or those with relatively simple native topology is well understood. Proteins, such as SH3 domain or Chymotrypsin Inhibitor 2, fold in an ostensibly two state manner although when examined using high spatial and temporal resolution it is found they too fold by multiple routes to the native state. For these proteins the yield of the native state is sufficiently large that their folding does not require the assistance of chaperones. 
However, from the perspective of assisted folding, it is more instructive to consider the folding of SPs whose folding landscapes are rugged containing many free energy minima (Fig. \ref{Landscapes_KPM}(a)) separated by sufficiently large barriers (several $k_BT$s)  that they cannot be overcome readily. Although the structures in the low energy minima could have considerable overlap with the folded state they are misfolded because they are likely to contain incorrect tertiary contacts and/or secondary structures. These are targets for recognition by molecular chaperones.  After an initial rapid compaction of the SPs or the ribozyme many of the molecules are trapped in one of several low free energy minima. 

The KPM, which explains the folding mechanisms of proteins succinctly,  follows immediately from the rugged folding landscape in Fig.\ref{Landscapes_KPM}(a) (see also Fig.\ref{3state}). 
According to KPM \cite{Guo95Biopolymers,Thirumalai97TCA}, a fraction of molecules $\Phi$ folds rapidly without being trapped in one of the low free energy minima. These are sometimes referred to as the fast track molecules for which, following an initial ``specific" collapse, folding to the native state is rapid \cite{Thirum95JPI}. Explicit simulations using lattice models \cite{Ziv09PCCP} have shown that the folding characteristics (dynamics of compaction and the increase in the fraction of native contacts as a function of time) of the fast track molecules are identical to sequences for which the folding landscape is simple with one dominant minimum. The remaining fraction ($1 -\Phi$) of molecules are trapped in an ensemble of low free energy structures because their initial collapse  produce structures containing interactions that are not present in the native state.    The resulting  misfolded structures have to overcome activation barriers in order to reach the folded state.  Thus, after the ensemble of unfolded molecules undergoes rapid collapse they partition to the native state at a rate $k_{IN}$ or transition to the misfolded ensemble at a rate $k_{IM}$.  The fraction of fast track molecules, referred to as the partition factor is associated with the rates in the 3-state cyclic model for chaperone-assisting folding depicted in Fig.\ref{3state} as, $\Phi = \frac{k_{IN}}{k_{IN} + k_{IM}}$. It is the value of $\Phi$, which depends on a number of extrinsic factors such as ionic strength, pH, and temperature that governs the need of a SP or the ribozyme for the chaperone machinery (see below). 

We  classify the misfolded structures  into slow folders and no folders,  depending on the magnitude of the activation free barriers separating them from the native state. The time scale for slow folders to reach the native state could range from milliseconds to several minutes whereas for no folders the transition to the native state could occur on time scales that exceed biologically relevant times. The effects of external conditions might be appreciated by noting that  ribulose bisphosphate carboxylase oxygenase (RUBISCO) behaves as a no folder at low ionic strength but becomes a slow folder at high ionic strength \cite{Schmidt94JBC}. Similarly, by increasing the temperature from about $(10 - 20)$ $^{\circ}$C to physiological temperature (37 $^{\circ}$C) both malate dehydrogenase and aspartate transaminase transition from being a slow to no folders \cite{Schmidt94JBC,Mattingly95JBC}. The no folders, with low $\Phi$, are prime candidates, which can fold with the aid of the complete chaperonin machinery.

\section{Experimental evidence for KPM} 
The KPM has been validated in a number of experiments. The value of $\Phi$ has been measured in ensemble and single molecule experiments. (i)  For example, Kiefhaber \cite{Kiefhaber95PNAS} showed, using interrupted folding (final folding condition is 0.6 M GdmCl, pH = 5.2 and $T=20^{\circ}$C), that $\Phi=$ 0.15 for hen egg white lysozyme (HEWL). The fast track molecules fold in about 50 ms. The remaining fraction are slow folders, which reach the native state on a time scale about 400 ms. By varying pH the value of  $\Phi$ was found to increase to about 25 \% \cite{Matagne1998JMB} while the time constants for the fast track molecules were roughly identical to the earlier study \cite{Kiefhaber95PNAS}. Because of the time for reaching the folded state by the molecules in the slow track is relatively small compared to biological times, it might be correctly concluded that folding of HEWL would not require the assistance of chaperones. (ii) The yield of the  folded RUBISCO obtained in the direct folding, under non-permissive conditions, inferred from chaperonin-mediated folding (see below) is extremely small and is only order of (2-5)\%. For both {\it Tetrahymena} and RUBISCO most of the molecules ought to be classified as no folders, which imply their folding requires molecular chaperones. (iii) An indirect estimate of $\Phi$ was first made using theory and experiments for {\it Tetrahymena} ribozyme. It was found \cite{Pan1997JMB} that $\Phi \approx 0.1$, which was subsequently confirmed in smFRET experiments \cite{Zhuang00Science}. These values were obtained at sufficiently high Mg$^{2+}$ concentration. At cellular Mg$^{2+}$ the value is expected to be much less. Introduction of a single point mutation (U273A) stabilizing the P3 pseudoknot helix was shown to increase $\Phi$ as high  to 0.8 \cite{Pan00JMB}, which shows that both sequence and external conditions determine the value of $\Phi$. For both {\it Tetrahymena} and RUBISCO most of the molecules ought to be classified as no folders, which imply their folding requires molecular chaperones.
In addition to the above mentioned experiments single molecule pulling experiments on several proteins (Tenacin, Fibronectin, T4Lysozyme, Calmodulin) using both Atomic Force Spectroscopy \cite{Li08PNAS,Stigler11Science}, and optimal tweezer techniques have established the validity of the KPM. 
\\

\subsection{Size and kinetic Constraints} 
Two constraints must be satisfied for GroEL-GroES assisted folding. First, pertains to the size of the SPs. The radius of gyration, $R_g$, of folded states of globular proteins is fairly accurately given by $R_g = 3N^{1/3}$\AA \cite{Dima04JPCB}. Small Angle X-Ray Scattering experiments on a few proteins have shown that the typical sizes of misfolded SPs  is about (5-10)\% larger than the folded states. This implies that the size of the RUBISCO monomer, with $N=491$, in the misfolded state is $\approx$ 32\AA.   The size of the expanded cavity, when both ATP and GroES are bound to GroEL,  is $\approx$ 185,000 \AA$^3$. If the cavity is approximated as a sphere the apparent radius would be 35 \AA, which implies that if RUBISCO is fully encapsulated in  the expanded cavity there would be room for about one layer of water molecules. Thus, GroEL can process SPs that contain $\lesssim$ 500 residues by fully encapsulating them. 

The second and a more important constraint is kinetic in nature. As argued before only a small fraction, $\Phi$ of the SPs, reaches the folded state rapidly without being kinetically trapped in one of the many metastable states. If the average rate for molecules that fold by the slow track  is $k_s$ then in order to prevent aggregation the pseudo first order binding rate, $k_B$, of the misfolded SP to bind to GroEL must greatly exceed $k_A$ where $k_A$ is a pseudo first order rate for SP aggregation. The kinetic constraint shows clearly that the efficacy of assisted folding depends on the concentrations of both the SPs and GroEL.   

\subsection{Allosteric Transitions in GroEL} 
Because the equilibrium  and non-equilibrium aspects in the spectacular allosteric transitions in GroEL have been recently reviewed \cite{Gruber16ChemRev,Thirumalai18PhilTrans,Lorimer18PhilTrans,Thirumalai19ChemRev},  we describe only briefly the key events that impact the nature of assisted folding.  Although the functional state of GroEL-GroES in the presence of SPs is the symmetric structure with the co-chaperonin bound to both the rings \cite{Sameshima10BiochemJ,Takei12JBC,Ye13PNAS,Yang13PNAS}, let us consider for illustration purposes only the hemicycle, thus allowing us to describe  events in one ring.  The $T$, $R$, and $R^{\prime\prime}$ are the three major allosteric states (Fig.\ref{transitions}).  
The misfolded SPs, with exposed hydrophobic residues, preferentially interact with the $T$ state, which has almost a continuous hydrophobic region  lining the mouth of the GroEL cavity. The presence of the hydrophobic region is due to the alignment of seven subunits that join several large hydrophobic residues in the two helices  (H and I) in the apical domain of each subunit. The $T \rightarrow R$ transition, resisted by the SP,  is triggered by ATP binding to the seven sites in the equatorial domain.  The rates of the  reversible $T \leftrightarrow R$ transition were first measured in pioneering studies by Yifrach and Horovitz \cite{Horovitz01JSB,Yifrach95Biochem} who also established an inverse relation, predicted using computations \cite{Betancourt99JMB},  between the extent of co-operativity in this transition and the folding rates of slow folding SPs \cite{Yifrach00PNAS}. Binding of GroES, which predominantly occurs only after ATP binds, drives GroEL to the so-called $R^{\prime}$ state, which is followed by an irreversible non-equilibrium transition to the $R^{\prime\prime}$ state after ATP hydrolysis. It is suspected there is little structural difference between the $R^{\prime}$, with ATP-bound, and the $R^{\prime\prime}$, containing  ADP and inorganic phosphate, states.  In both these transitions strain due to ATP binding and hydrolysis at the catalytic site propagates through a network of inter-residue contacts \cite{Tehver09JMB}, thus inducing  large scale conformational changes. That such changes must occur during the reaction cycle of GroEL is already evident by comparing  the static crystal structures in different allosteric states, such as the $T$ and $R^{\prime\prime}$ states \cite{Xu97Nature}. Release of ADP and the inorganic phosphate from the $R^{\prime\prime}$ state resets the machine back to the taut state from which a new cycle can begin. The allosteric transitions that GroEL undergoes during the catalytic cycle is intimately related to its function (see below).
As we discuss later, it is not sufficient to deal with the catalytic cycle in a single ring because under load it is the symmetric football-like structure that is the functional state.
\\

\subsection{Iterative Annealing Mechanism integrates GroEL Allostery and assisted SP folding} 
\begin{figure*}[t]
\centering
\includegraphics[width=0.9\textwidth]{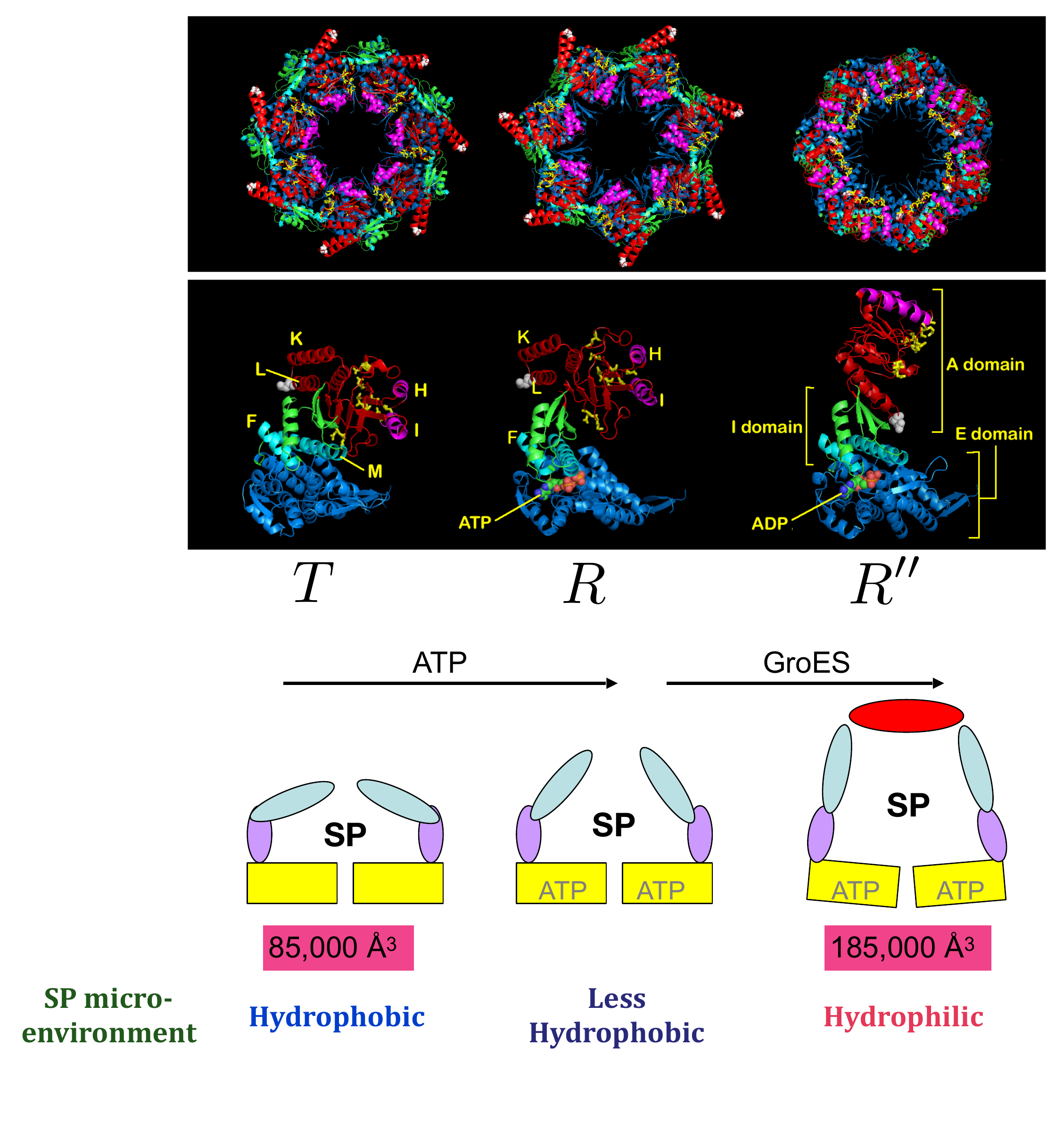}
\caption{Allosteric states in GroEL. The $T \rightarrow R$ transition is driven by ATP, and subsequent binding of GroES and ATP hydrolysis results in the $R \rightarrow R^{\prime\prime}$. 
As a result of transition from $T$ to $R^{\prime\prime}$, 
the volume of cavity expands from 85,000 to 185,000 \AA$^3$, and the SP experiences the change in microenvironment from hydrophobic in the $T$ state to hydrophilic in the $R^{\prime\prime}$ state. 
}
\label{transitions}
\end{figure*}

\begin{figure*}[t]
\centering
\includegraphics[width=0.9\textwidth]{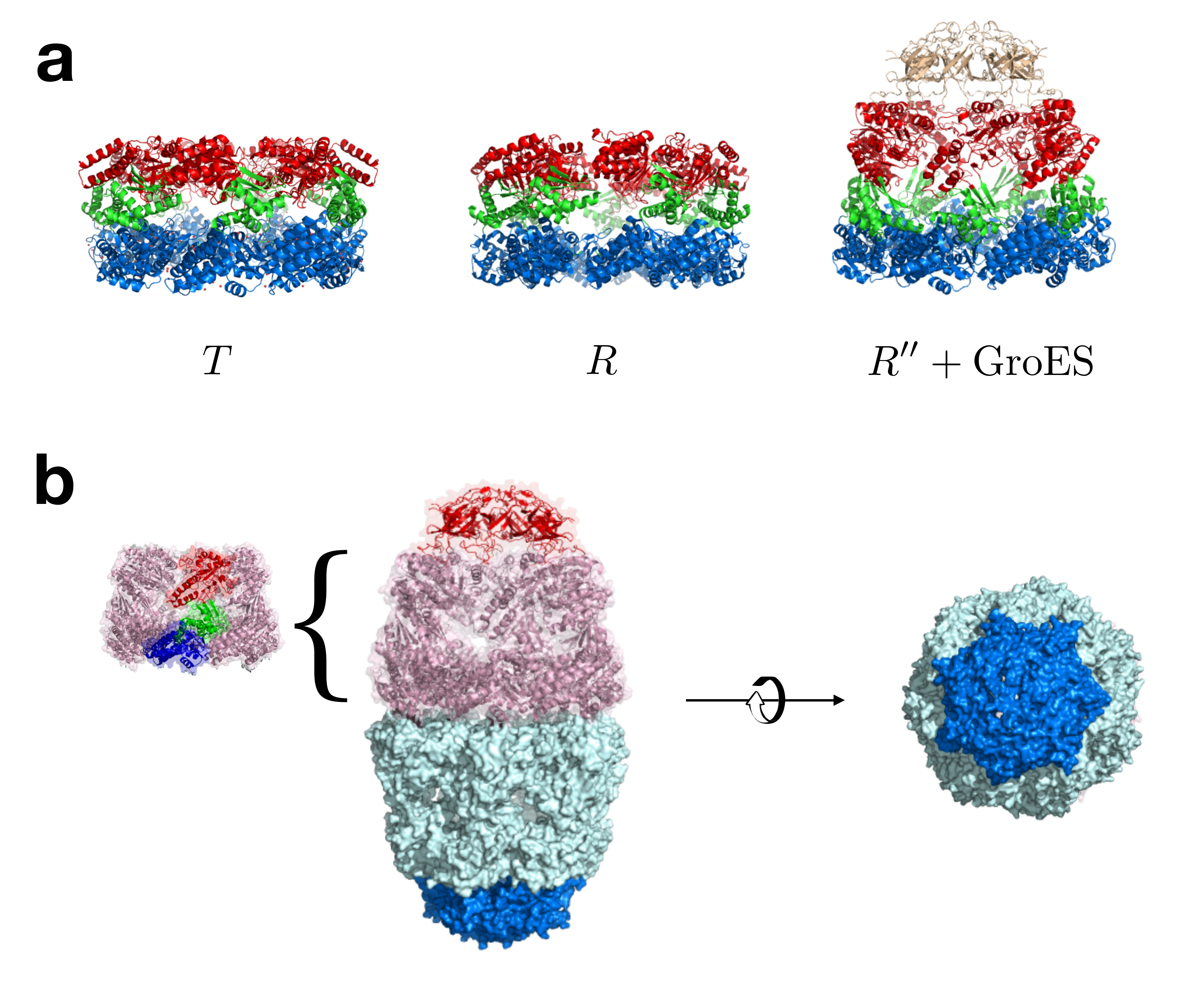}
\caption{Structures of the chaperonin GroEL-GroES molecular. 
{\bf a}. Structures of GroEL in $T$, $R$, and $R^{\prime\prime}$ states (PDB codes: 1OEL, 2C7E, 1AON). 
Apical, intermediate, and equatorial domains are colored in red, green, and blue, respectively. 
In $R^{\prime\prime}$ state, GroES is bound on top of the apical domain of GroEL ring structure. 
{\bf b}. Football like structure of GroEL$-$GroES complex (PDB code : 4PKN) is the functional state that is populated in the presence of substrate proteins. A view from the bottom highlights the structure with 7-fold symmetry.}
\label{structure}
\end{figure*}

\begin{figure*}[t]
\centering
\includegraphics[width=0.9\textwidth]{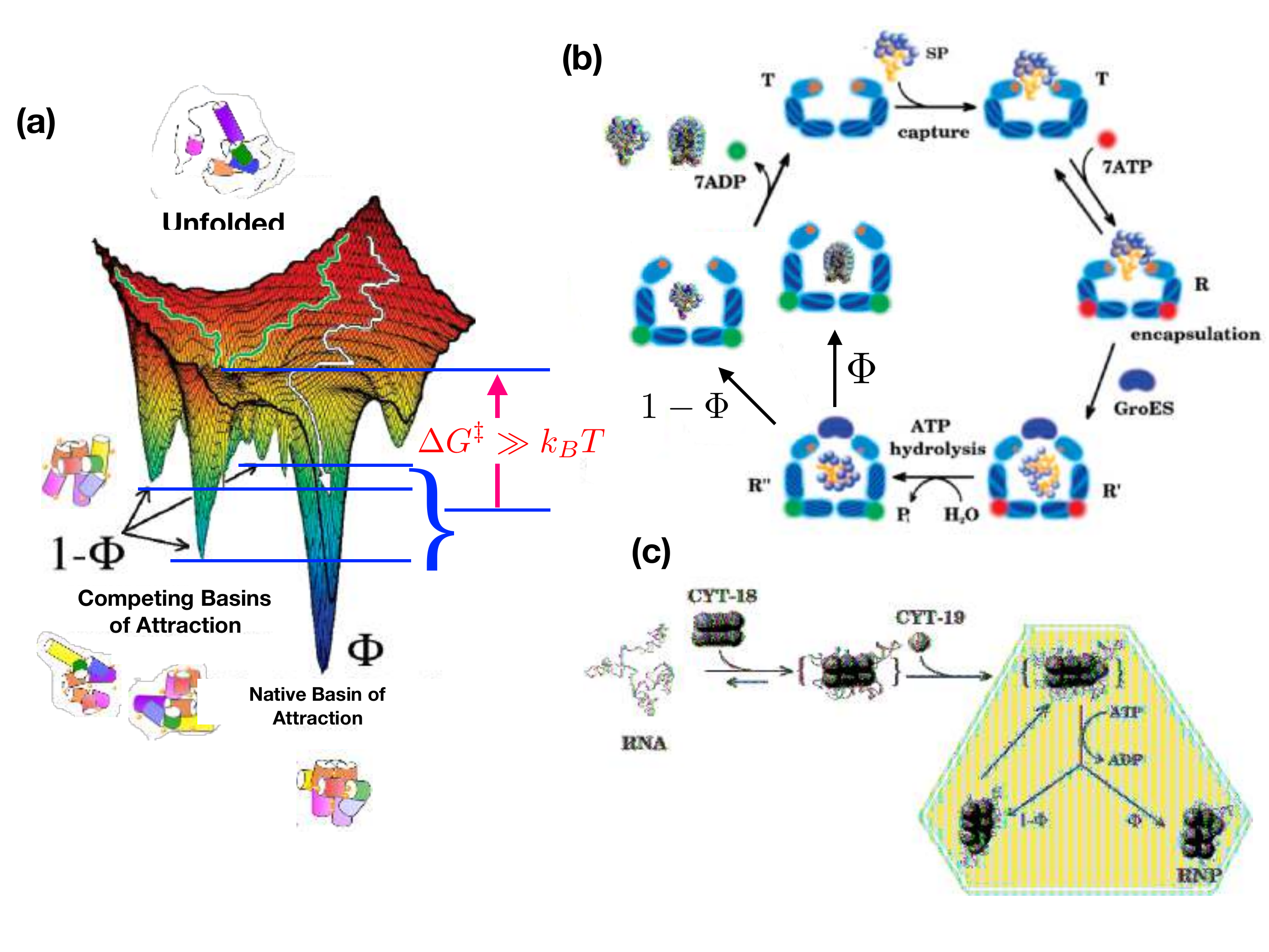}
\caption{Kinetic partitioning mechanism (KPM) on rugged energy landscapes and chaperones.  
(a) Rugged folding landscape illustrating the native (NBA) and competing basins of attraction (CBA). 
In the absence of chaperones, a fraction $\Phi$ of the unfolded state ensemble folds into the NBA and $1-\Phi$ of the ensemble collapses to CBA. 
(b) IAM for GroEL-GroES showing the coupling between allosteric transitions and SP folding. The figure clearly illustrates that partitioning to native state, with probability $\Phi$, and mifolding to a metastable state, with probability ($1 - \Phi$) occurs rapidly within the cavity during the two second life time of the encapsulated state. Although not shown explicitly, the functioning state is the symmetric complex (see Fig. \ref{structure}). 
(c) IAM for RNA chaperone (CYT-18/CYT-19) acting on RNA. 
Both the processes shown in (b) and (c) are energy consuming processes associated with ATP hydrolysis. 
}
\label{Landscapes_KPM}
\end{figure*}

\begin{figure*}[t]
\centering
\includegraphics[width=1\textwidth]{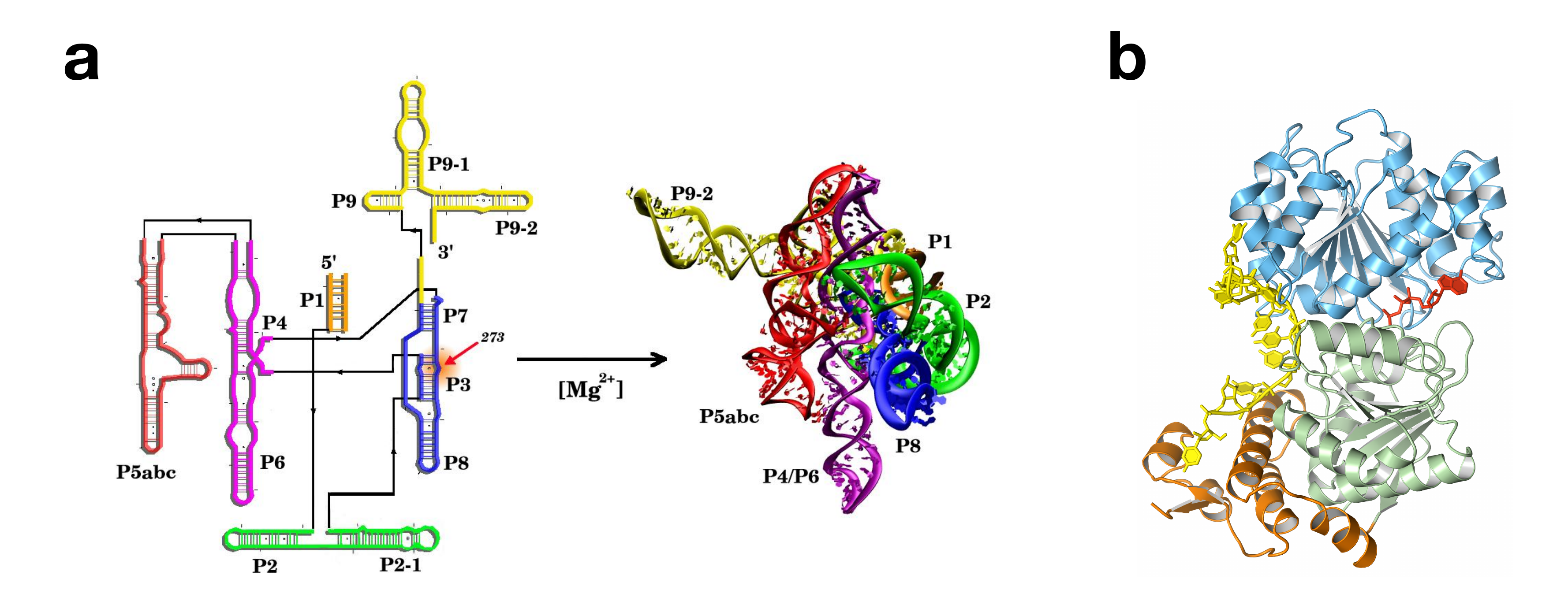}
\caption{{\bf a}. Folding of \emph{T}. ribozyme from its secondary structure to three dimensional native state. 
{\bf b}. Structure of yeast analogue of CYT-19.}
\label{RibozymeFig}
\end{figure*}

\begin{figure*}[t]
\centering
\includegraphics[width=0.7\textwidth]{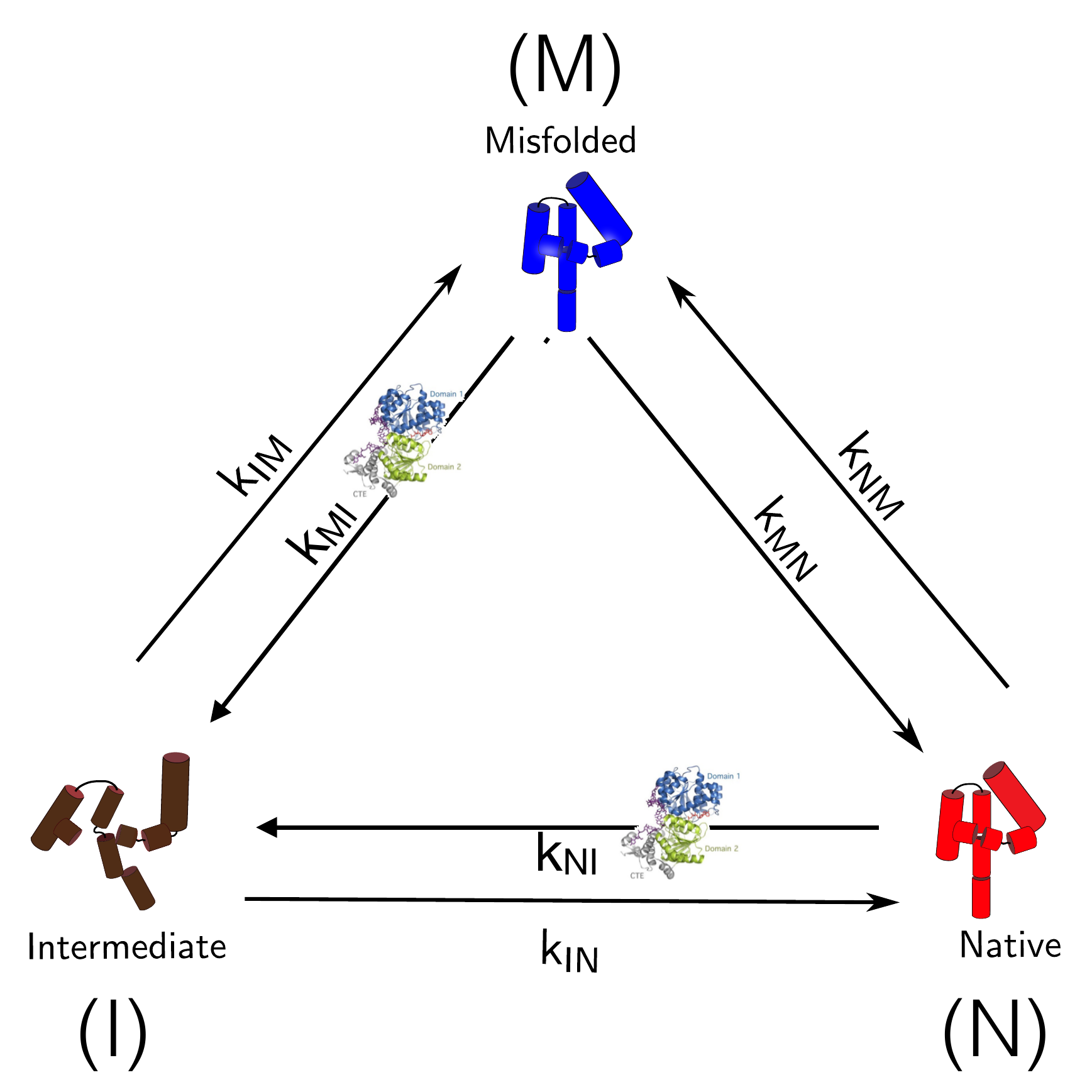}
\caption{The model for a unified description of chaperone-assisted folding. 
Tetrahymena ribozyme and CYT-19 (green) are employed for illustration purposes. 
The model shows the ribozyme in the I (brown), N (red), and M (blue) states and ribozyme-CYT-19. 
The GroEL-associated folding can similarly be accounted for by replacing CYT-19 with the chaperonin machinery.
The figure was adapted from Ref.\cite{chakrabarti2017PNAS}.
}
\label{3state}
\end{figure*}

The importance of GroEL allostery in assisted folding can be appreciated by understanding the interaction of the SP with the GroEL-GroES system in different allosteric states. The changes in the SP-GroEL interaction occur in three stages corresponding to the allosteric transitions between the three major allosteric states (see Fig.\ref{transitions}). (i) The continuous lining of the hydrophobic residues in the $T$ state ensnares a misfolded SP with exposed hydrophobic residues.     At this stage in the catalytic cycle the SP is predominantly in a hydrophobic environment, resulting in the formation of a SP-GroEL complex that is stable but not hyper stable so that the SP can be dislodged in to the cavity upon GroES binding \cite{ThirumalaiBookDoniach,Orland97JP}. (ii)The dynamics of the $T \leftrightarrow R$ transition, upon ATP binding, reveals that there is a downward tilt in two helices near the E domain that closes off the ATP binding sites, and which is followed by multiple salt bridge disruption (within a subunit) and formation of new ones across the adjacent subunits \cite{Hyeon06PNAS}. As  these events unfold cooperatively, the stability of the initial SP-GroES complex decreases. More importantly,  the adjacent subunits start to move apart, which imparts  a moderate force that is large enough to at least partially unfold the SPs \cite{ThirumARPC01,Corsepius13PNAS}.   (iii) Both GroES and SP bind to the same sites, which are located in the crevices of helices H and I in the apical domain. 
Thus, when GroES binds, displacing the SP into the expanded central cavity, there are major structural changes in the GroEL cavity with profound consequences for the annealing mechanism. 
Only 3--4 of the 7 SP binding sites are needed to capture the SP, leaving 3--4 sites available for binding of the mobile loops of GroES. This ensures that the subsequent displacement of the SP occurs vectorially into the central cavity of GroEL.
First, there is a significant conformational change in the A domain, which undergoes a rotation and twist motion.  
Each subunit results in the two helices (K and L) in each subunit undergo an outside-in movement (Fig.\ref{transitions}). 
As a result, polar and charged residues, which are solvent exposed in the $T$ state, line the inside of the GroEL cavity. This in turn creates a polar microenvironment for the SP (Fig.\ref{transitions}).  
Second, these large scale conformational changes are facilitated by the formation of several inter subunit salt bridges and disruption of intra subunit salt bridges \cite{Hyeon06PNAS}. 

From the perspective of SP, there are major consequences that occur as a result of the allosteric transitions in GroEL. First, by breaking a number of salt bridges  the volume of the central cavity increases two fold (85,000 \AA$^3$$\rightarrow$185,000 \AA$^3$). In such a large central cavity, enough to fully accommodate a compact protein with $\approx$ 500 residues, folding to the native state could occur if given sufficient time as is the case in the SR1 mutant. 
But in the wild type the residence time of the encapsulated SP is very short (see below). Second, and most importantly, the SP-GroEL interaction changes drastically during the catalytic cycle. In the  $T$ state,  SP-GroEL complex is (marginally) stabilized predominantly by hydrophobic interactions. However, during the subsequent ATP-consuming and irreversible step  $R \rightarrow R^{\prime\prime}$ transition the microenvironment of the SP is largely polar (see the discussion in the previous paragraph). Thus, during a single catalytic cycle, that is replicated in both the rings, the microenvironment of the SP changes from being hydrophobic to polar. We note parenthetically that even during the $T \rightarrow R$ transition there is a change in the SP-GroEL interactions, which explains the observations that GroEL can assist of the folding of certain SPs (non stringent substrates) even in the absence of GroES. The  annealing capacity of GroEL is intimately related to the changes in the SP-GroEL interactions that occur during each catalytic cycle. Hence, the function of the GroEL-GroES system cannot be understood without considering the complex allosteric transitions that occur due to ATP and GroES binding.  As a result of these transitions,  the SP is placed stochastically from one region in the folding landscape, in which the misfolded SP is trapped, to another region from which it could undergo kinetic partitioning with small probability to the folded state or be trapped in another misfolded state. The cycle of hydrophobic to polar change is repeated in each catalytic cycle, and hence the GroEL-GroES system iteratively anneals the misfolded SP enabling it to fold to the native state. Because this process is purely stochastic, GroEL plays {\bf no role} in guiding the protein to the folded state nor does it sense the architecture or any characteristics of the folded state. In other words, the information for protein self-assembly is fully encoded in the amino acid sequence as articulated by Anfinsen \cite{Anfinsen73Science}. GroEL merely alters the conformation of the SP stochastically as it undergoes the reaction cycle, enabling the SP to explore different regions of the folding landscape. In this sense the action of GroEL is analogous to simulated annealing used in optimization problems \cite{Kirkpatrick83Science} although the latter is a more recent realization of an evolutionary event that took place millions of years ago.

\subsection{Theory underlying the IAM} 
The physical picture of the IAM described above can be formulated mathematically to quantitatively describe the kinetics of chaperonin-assisted folding of stringent {\it in vitro} substrate proteins \cite{Tehver08JMB}. According to theory underlying IAM (see Fig. \ref{Landscapes_KPM}), in each cycle the SP folds by the KPM, as the microenvironment for the SP changes as GroEL undergoes the reaction cycle.   Thus, with each round of folding the fraction of folded molecules is $\Phi$, and the remaining fraction gets trapped in one of the many misfolded structures. After $n$ such cycles (or iterations) the yield of the native state is,
\begin{equation}
\Psi = \Lambda_{ss} \left[1- (1 - \Phi)^n\right]
\label{yield}
\end{equation}
where $\Lambda_{ss}$ is the steady state yield. The mathematical model accounts for all the available experimental data, and shows that for for RUBISCO the partition factor $\Phi \approx 0.02$, which means that only about 2\% of the SP reaches the folded state in each cycle.  From this finding we could surmise that the GroEL chaperonin is an inefficient machine, which consumes ATP lavishly and yet the yield of the folded protein per cycle is small. A prediction of the IAM is that GroEL should reset to the starting $T$ state as rapidly as possible in the presence of SPs. By rapidly resetting to the $T$ state the number of interactions can be maximized, which would maximize the yield of the folded state for a specified amount of time \cite{chakrabarti2017PNAS}. Indeed, this is the case, which we delve into detail below. \\

\subsection{Rate of $R^{\prime\prime} \rightarrow T$  transition is a maximum for the wild-type (WT) GroEL} 
A clear implication of the IAM is that rapid turnover of the catalytic cycle would produce the maximum yield of the native state in a given time. Examination of the reaction cycle shows that the rate determining step (resetting of the machine) should correspond to release of ADP and the inorganic phosphate. In other words, maximization of the rate, $k_{R^{\prime\prime} \rightarrow T}$ returns GroEL to the acceptor state for processing a new SP. In order to illustrate that this is indeed the case, we first extracted  the rates of the allosteric transitions by fitting  the solutions of the kinetic equations \cite{Tehver08JMB} by simultaneously fitting the experimental  data for assisted folding at various GroEL concentrations. For this purpose, we used the data for RUBISCO for which the yield of the folded state as a function of eight values of the GroEL concentration are available \cite{Todd96PNAS}. 
The excellent fits at various GroEL concentrations (Fig.\ref{Tehver2a08JMB}), with a fixed initial concentration of Rubisco, were used to extract $\Phi$. We find that $\Phi \approx 0.02$, which means that only about 2\% of the SP reaches the folded state in each catalytic cycle. 

Armed with the rates that describe the allosteric transitions, we  used the IAM theory based to analyze experimental data on the folding of other SPs. Because the reversible transition ATP-induced $T \leftrightarrow R$ transition occurs at equilibrium even in the absence of SP \cite{Yifrach95Biochem} it is reasonable to assume that they are relatively insensitive to the nature of the SP. Indeed, the extracted values of the $T \leftrightarrow R$  rates using the RUBISCO data (see Table 1 in \cite{Tehver08JMB}) are very similar to measurements made in the absence of SP \cite{Yifrach95Biochem}. This leaves  the rate $k_{R^{\prime\prime} \rightarrow T}$ that results in the resetting the machine after ATP hydrolysis to the taut ($T$) state as the most important factor in determining the efficiency of GroEL or its mutants. Thus,  maximizing $k_{R^{\prime\prime} \rightarrow T}$ should result in optimizing the native state yield at a fixed time. This most significant prediction of IAM can be quantitatively demonstrated by analyzing the data reported by Lund and coworkers \cite{Sun2003JMB}. They measured the activity, which we assume is proportional to the yield of the folded state, as a function of time for GroEL and five mutants including SR1. The two SPs used in these studies were mitochondrial Malate Dehydrogenase (mtMDH) and citrate synthase (CS). The results, reproduced in Fig.\ref{mtMDH_CS} shows that $P_N^{SS}$ and indeed the yield at any time  is largest for the WT and is least for the SR1 mutant from which GroES disassociates in $\approx$ 300 minutes. The curves in Fig.\ref{mtMDH_CS} were calculated by adjusting just {\it one parameter}, the rate $k_{R^{\prime\prime} \rightarrow T}$ while keeping the rates for other allosteric rates fixed at the values extracted by analyzing the RUBISCO data.   The IAM predictions are in {\bf quantitative} agreement with experiments for both the proteins and for GroEL and its mutants. The value of $k_{R^{\prime\prime} \rightarrow T} \approx$ 60 s$^{-1}$ (one second) is largest the WT GroEL. This implies, as predicted by IAM, that GroEL catalytic cycle is greatly accelerated when SP is present, a point that requires further elaboration.  
\\

\section{Generalized IAM for RNA chaperones}
Compared to GroEL-GroES chaperonin, details of the catalytic cycle of CYT-19 are not known. Consequently, it is not possible to link the structural transitions that occur during the CYT-19 assisted folding of the misfolded ribozymes, as we did for the GroEL-GroES machinery.  We should note that the structures and biophysical studies of the DEAD-box protein Mss116p,  {\it Saccharomyces
cerevisiae} analogue of CYT-19, showed the expected helicase activity, resulting in the disruption of the structure of the misfolded ribozyme. These studies and the still undetermined ATPase cycle could be used in the future to provide a molecular basis of the IAM for CYT-19 assisted folding. Nevertheless, the mathematical formulation of the IAM theory could be adopted to investigate the interesting experimental findings by Bhaskaran and Russell \cite{Bhaskaran07Nature}.  

The most significant experimental findings of CYT-19 assisted folding of ribozymes are:
(i) When incubated in CYT-19 under somewhat destabilizing conditions ([Mg$^{2+}$]$<$ 2 mM), ribozymes show a low cleavage activity. 
(ii) Regardless of the initial population, the native and the misfolded ribozyme reach a steady state value for the folded ribozyme fraction, which is not unity ($P^{SS}_N\neq 1$). 
(iii) The deactivation of the ribozyme function was observed at longer pre-incubation times in CYT-19. Deactivation of native ribozyme was also observed at higher CYT-19 concentration. 
Taken together, these observations imply that CYT-19 destabilizes the native as well as the misfolded ribozyme. The finding that CYT-19 interacts with the fully folded ribozyme is in stark contrast with GroEL, which does not interact with the folded states of proteins.  
In light of the experimental observations the  IAM theory has to be generalized (see Fig.\ref{Generalized_IAM}). The results in this study inspired us to generalize the IAM theory using the master equation \cite{Hyeon2013JCP}. More recently, we proposed a simpler version that describes the functions of GroEL and RNA chaperones on equal footing \cite{chakrabarti2017PNAS}. The resulting theory, which gives rise a complicated expression for the folded state of P5a variant of the {\it Tetrahymena} sketched below,  provided a quantitative agreement (Fig.\ref{ShaonFig5ab}) of  the experimental data \cite{Bhaskaran07Nature}.  

The KPM description of  ribozyme folding  \cite{Pan1997JMB,Thirumalai97TCA} shows that  upon increasing the  Mg$^{2+}$ concentration 
a fraction of the initial unfolded population, $\Phi$, folds to the native state and the remaining fraction, $M_1=1-\Phi$, collapses to one of many misfolded states. Consider the fate of the misfolded states, with population, $M_1$, as they interact with CYT-19.
In the presence of the RNA chaperone, a fraction $\Phi$ of $M_1$  reaches the native state ($\Phi(1-\Phi)$) and $1-\Phi$ of $M_1$ to one of the misfolded states ($(1-\Phi)^2$).  Because CYT-19 also acts on the native state we also have to consider the fate of the folded ribozyme, as it interacts with CYT-19 (see Fig.\ref{Generalized_IAM}). Let a fraction $\kappa$ denote the fraction of the initially folded ribozyme reach the misfolded state (bottom right circle in Fig.\ref{Generalized_IAM}) while the $1-\kappa$ remain in the  native state (top right circle in Fig.\ref{Generalized_IAM}). In the subsequent round,   
Out of $\kappa \Phi$, $\kappa\Phi^2$ of them goes to native and $\kappa\Phi(1-\Phi)$ reaches the misfolded state. 
Therefore, $M_2=\kappa\Phi(1-\Phi)+(1-\Phi)^2$ is the total  of the misfolded ribozyme in the second round of IAM, which accounts for accumulation from both the folded and misfolded states in the first round. In order to obtain an expression for the yields of both the folded and misfolded states of the ribozyme the branching process from both the accumulated folded and misfolded states of the ribozyme in the previous round has to be taken into account.
A recursion relation for this iterative process may be written down, such that the amount of misfolded state at the $n$-th round is the sum of $M_{n-1}\times (1-\Phi)$ from the misfolded ensemble, and $\kappa(1-M_{n-1})(1-\Phi)$ from the native ensemble. 
In short, $M_n=M_{n-1}(1-\Phi)+\kappa(1-M_{n-1})(1-\Phi)$ (see Fig.\ref{Generalized_IAM}). 
As a result, the total yield of native state in the $N$-th round of annealing process ($\Psi_N=1-M_N$) can be calculated in order to obtain yield of the native ribozyme from the generalized version of IAM, 
\begin{align}
\Psi_N=\Phi\frac{1-(1-\kappa)^N(1-\Phi)^N}{\kappa+(1-\kappa)\Phi}, 
\end{align}
and the steady state solution ($N\rightarrow\infty$) 
is 
\begin{align}
\Psi_{\infty}=\frac{\Phi}{\kappa+(1-\kappa)\Phi}. 
\label{Psi_inf}
\end{align}
For $\kappa=0$, corresponding to the situation that the RNA chaperone does not recognize the native state , the yield in the $N$-th round is identical to the conventional IAM expression. 
For $\kappa=1$ in which RNA chaperone recognizes the native state  equally as well as misfolded states, there would be no gain in the native yield by the action of RNA chaperone. 

The action of chaperones on substrate RNA can be mapped onto 3-state kinetic model of RNA with transitions between the native ($N$), misfolded ($M$), and intermediate states ($I$). 
When the partition factor $\Phi$ in terms of the rate constants, $\Phi=k_{IN}/(k_{IN}+k_{IM})$, is plugged into Eq.\ref{Psi_inf}, $\Psi_{\infty}$ is 
\begin{align}
\Psi_{\infty}=\frac{k_{IN}}{\kappa k_{IM}+k_{IN}}. 
\label{Psi_infty}
\end{align}
In addition, the expression for the steady state value of fraction native ($P_N^{SS}$), which is equivalent to $\Psi_{\infty}$, can be obtained using 3-state kinetic model (Fig.\ref{3state}) under the following  conditions: (i) $k_{NM}$, $k_{MN}\ll k_{IN}$, $k_{IM}$, $k_{NI}$, $k_{MI}$, and (ii) $k_{NI}\ll k_{IN}$. With these assumptions  we find that,  
\begin{align}
P_N^{SS}
\approx \frac{k_{IN}}
{\left(\frac{k_{NI}}{k_{MI}}\right)k_{IM}+k_{IN}}. 
\label{P_N^ss}
\end{align}
Therefore, comparison between Eq.\ref{Psi_infty} and Eq.\ref{P_N^ss} gives 
\begin{align}
\kappa=\frac{k_{NI}([C],[T])}{k_{MI}([C],[T])}
\end{align}
where the dependence of unfolding rates $k_{NI}$ and $k_{MI}$ on chaperone and ATP concentration is made explicit. 
It turns out that $\kappa$, defined as the unfolding efficiency of chaperone for the native state with reference to the misfolded ensemble, is effectively the ratio between chaperone-induced unfolding rate from the native and misfolded state. A sketch of the native state as a function of $\kappa$, which depends both on the chaperone and ATP concentration, is given in Fig.\ref{Fig3Shaon17PNAS}. 

\section{Discussion}

\subsection{What do chaperones optimize?} 
\begin{figure*}[t]
\centering
\includegraphics[width=0.7\textwidth]{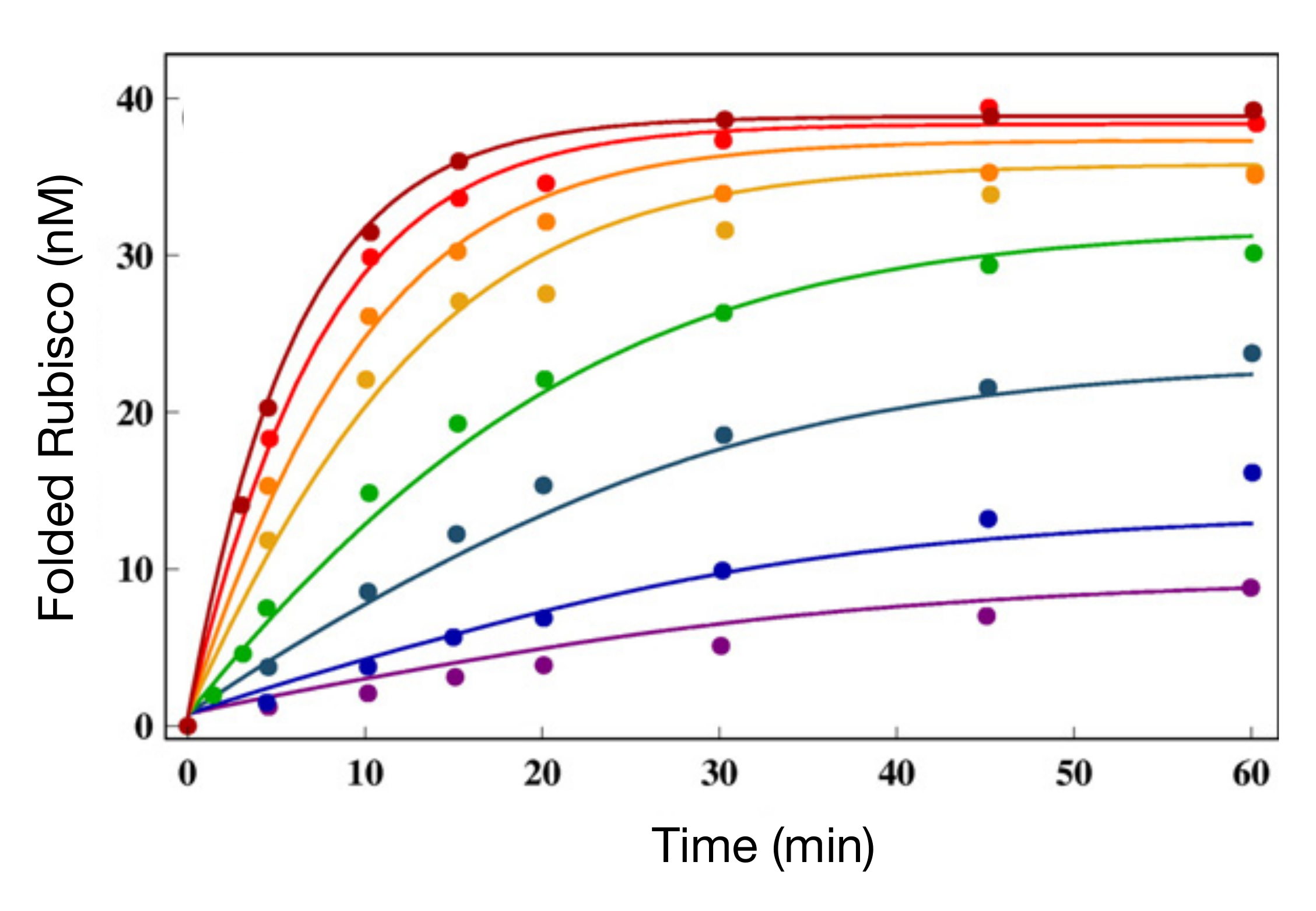}
\caption{Native-state yield of Rubisco as a function of time at different GroEL concentrations. 
The chaperonin concentrations for the curves from bottom to top are 1 nM, 2 nM, 5 nM, 10 nM, 20 nM, 30 nM, 50 nM, and 100 nM.
The lines are the fits to the experimental data from Ref. \cite{Todd96PNAS} using the kinetic model. 
The figure was adapted from Ref.\cite{Tehver08JMB}.  
}
\label{Tehver2a08JMB}
\end{figure*}

\begin{figure*}[t]
\centering
\includegraphics[width=0.6\textwidth]{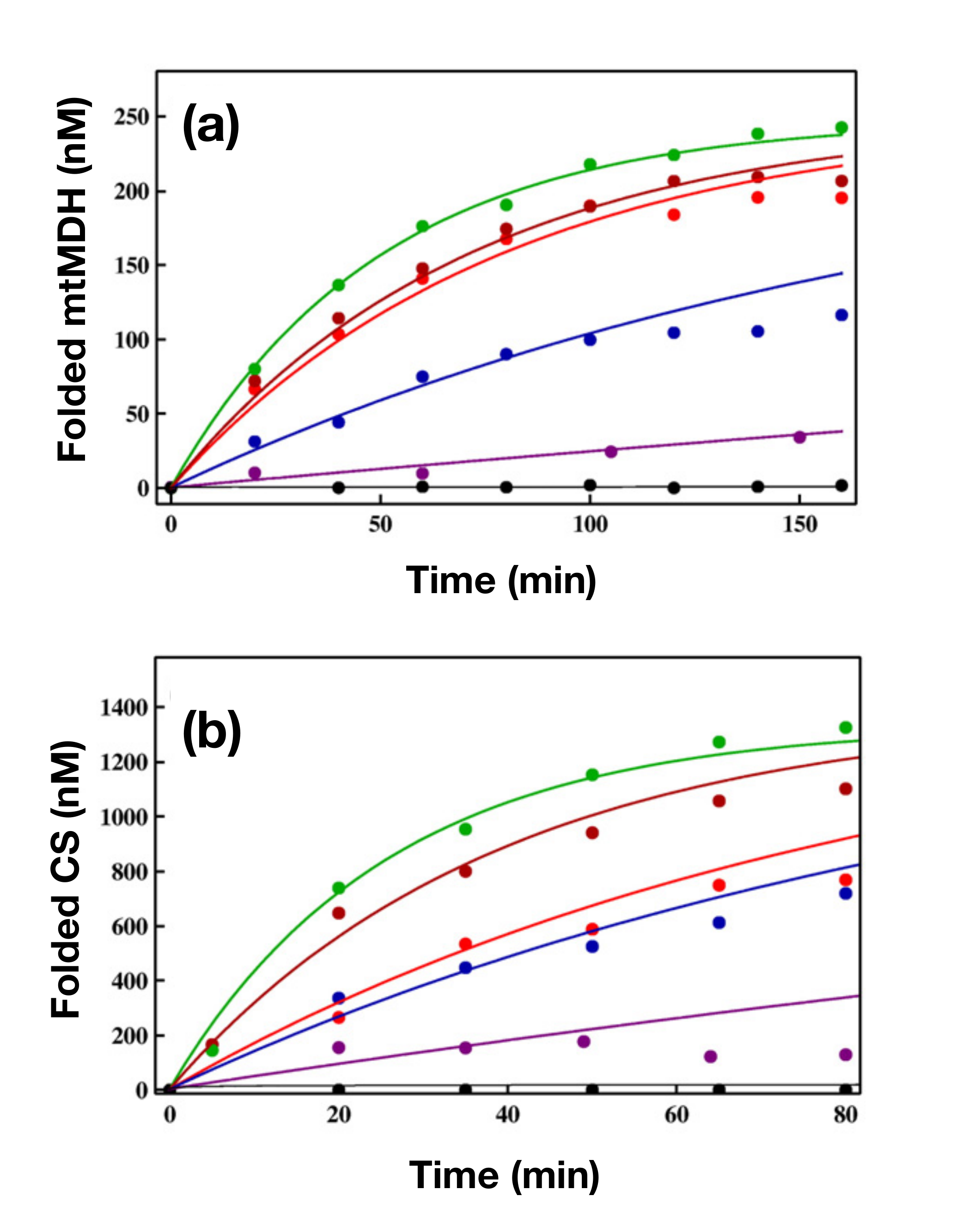}
\caption{Yield of SPs as a function of time. 
(a) The data points are taken from the experiment for folding of mtMDH. 
The lines are the fits using the kinetic model developed in \cite{Tehver08JMB}.
The black line is for spontaneous folding. Assisted folding in the presence of GroES and SR1 (purple), SR-T522I (blue), SR-A399T (red), and SR- D115N (dark red) was used to assess the efficiencies of these three single-ring chaperonins relative to GroEL (green). (b) The same as (a) except the SP is CS. 
}
\label{mtMDH_CS}
\end{figure*}

\begin{figure*}[t]
\centering
\includegraphics[width=0.9\textwidth]{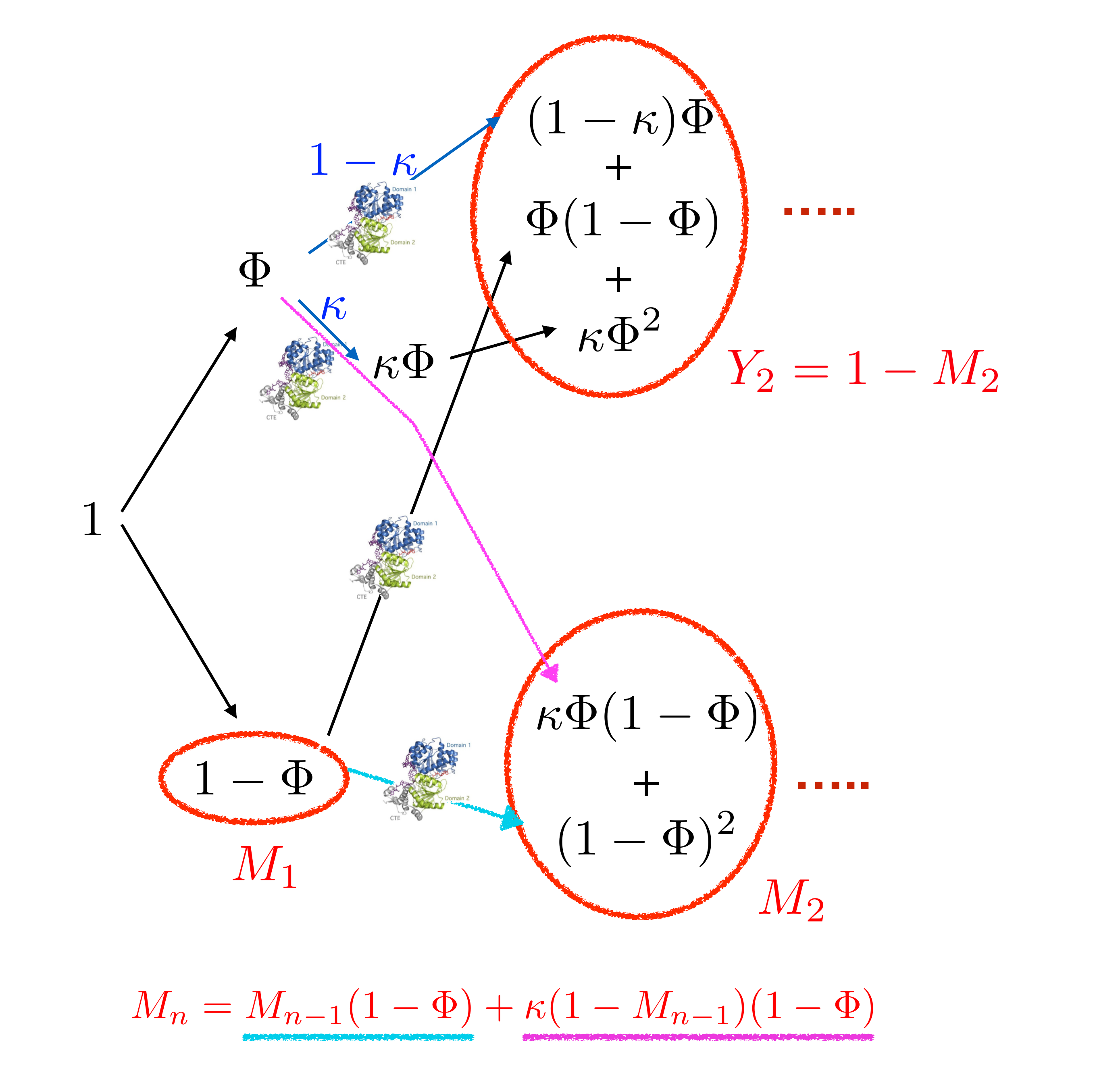}
\caption{Schematic of the generalized IAM of chaperone-assisted substrate folding. 
Depicted are the logical steps in a branching process  that leads to the recursion relation for the total yield of the misfolded state after $n$-th annealing process ($M_n$), given at the bottom. 
$Y_i(=1-M_i)$ and $M_i$ are respectively the yield of native and misfolded states from the $i^{th}$ iteration. 
$\Phi$ is the kinetic partitioning factor. 
}
\label{Generalized_IAM}
\end{figure*}

\begin{figure*}[t]
\centering
\includegraphics[width=0.9\textwidth]{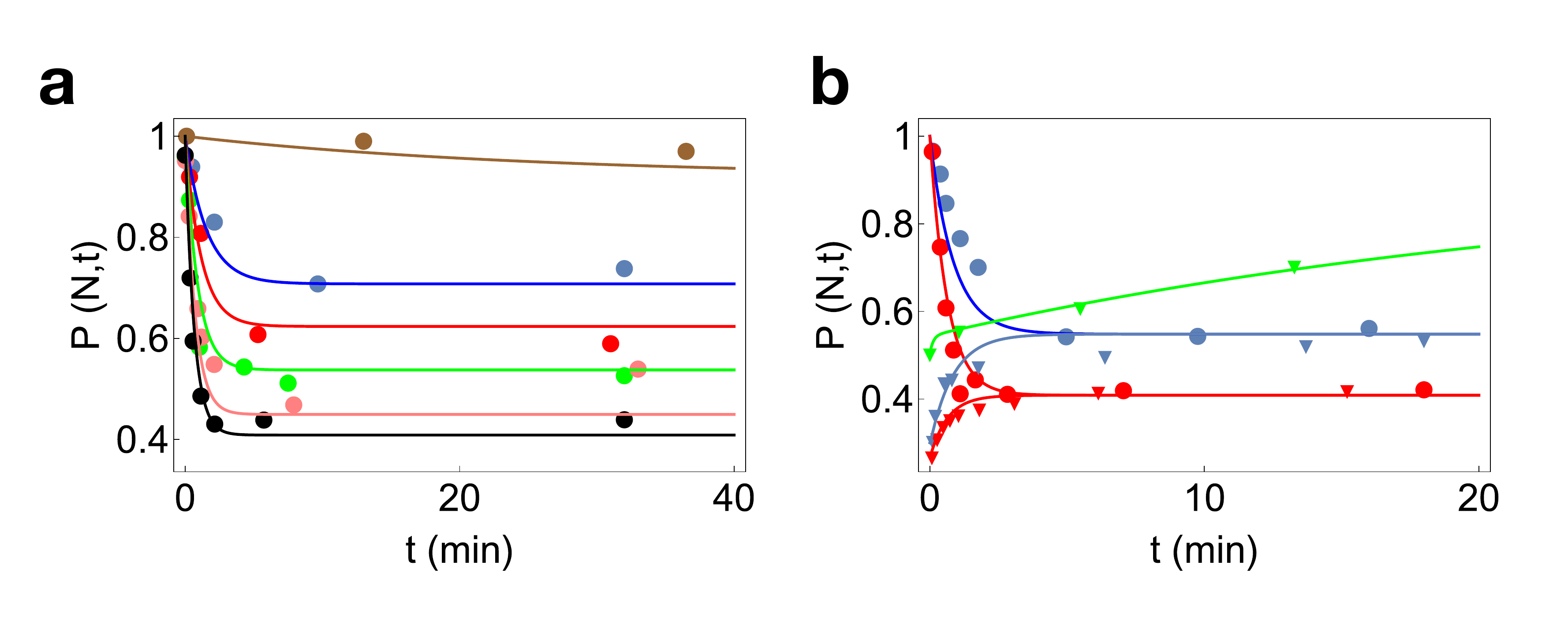}
\caption{Analysis of folding of the P5a variant of \emph{T}. ribozyme in the presence of CYT-19.  
(a) CYT-19 (1 $\mu$M)-induced kinetics of the native P5a variant ribozyme in 5 mM Mg$^{2+}$ at various ATP concentrations: no ATP (brown), 100 $\mu$M ATP (blue), 200 $\mu$M ATP (red), 400 $\mu$M ATP (green), 1 mM ATP (pink), and 2 mM ATP (black). (b) Kinetics of P5a variant folding for different CYT-19 concentrations. 
Starting conditions were native (circles) or misfolded (triangles) P5a variants. 
CYT-19s are 0.5 $\mu$M (blue) and 1 $\mu$M (red). 
The curve in green is for a mixture of native and misfolded P5a variant ribozymes when proteinase K is added to inactivate CYT-19. The figure was adapted from Ref.\cite{chakrabarti2017PNAS}.}
\label{ShaonFig5ab}
\end{figure*}

\begin{figure*}[t]
\centering
\includegraphics[width=0.6\textwidth]{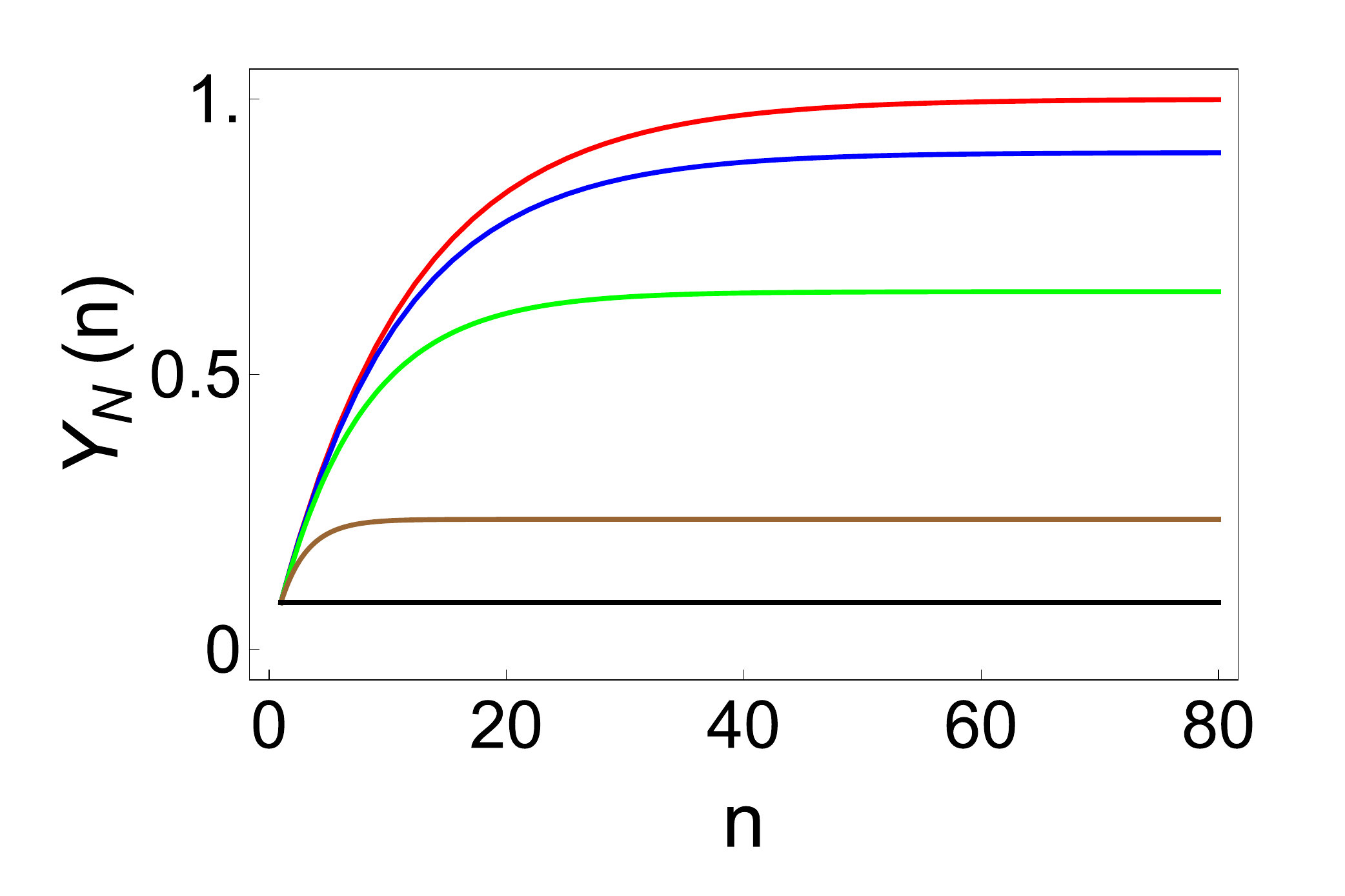}
\caption{The effect of varying $\kappa$ on the yield of the N state. . 
Shown is the plot of the native yield, $Y_N(n)$ as a function of number of cycles $n$ for varying $\kappa$ values:  
$\kappa = 0$ (red), $\kappa = 0.01$ (blue), $\kappa = 0.05$ (green), $\kappa = 0.3$ (brown), and $\kappa = 1.0$ (black).
The figure was adapted from Ref.\cite{chakrabarti2017PNAS}.}
\label{Fig3Shaon17PNAS}
\end{figure*}

\begin{figure*}[t]
\centering
\includegraphics[width=0.9\textwidth]{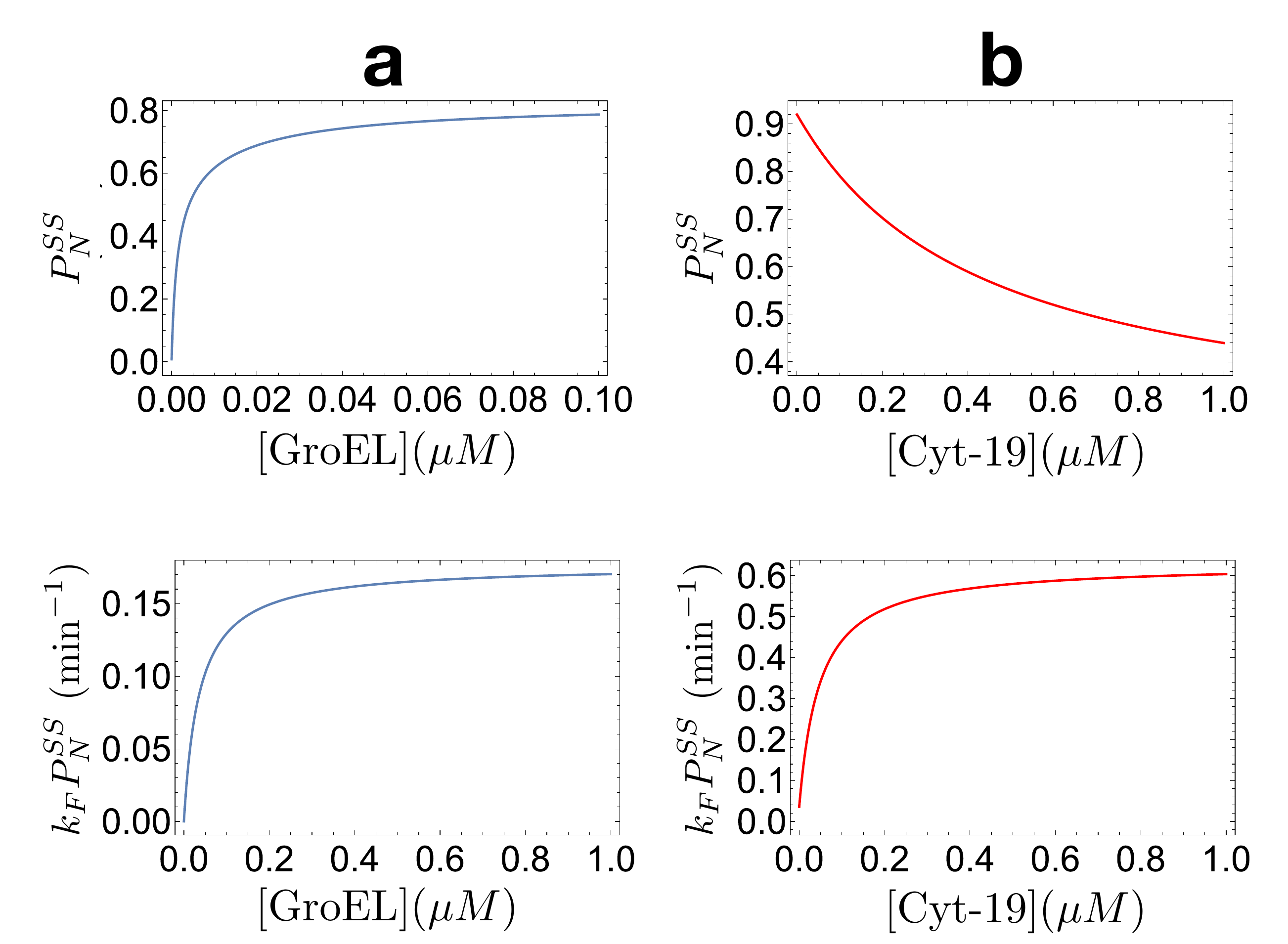}
\caption{Maximization of the finite-time yield by iterative annealing. {\bf a} GroEL and {\bf b} Cyt-19.  
(Top panels) Steady-state yield of the folded Rubisco ({\bf a}) and ribozyme ({\bf b}), 
as a function of chaperone concentration. 
(Bottom panels) Yield per unit time $\Delta_{NE} = k_F P_N^{SS}$ for Rubisco ({\bf a}) and ribozyme ({\bf b}), as a chaperone concentration. 
For all of the curves, ATP concentration was set to 1 mM. 
The figure was adapted from Ref.\cite{chakrabarti2017PNAS}. 
}
\label{Fig. 7Shaon17PNAS}
\end{figure*}

The question of what quantity a biological machines optimize subject to the constant of available free energy does not have a general answer. However, in the rare case of chaperones a plausible answer has been recently proposed, which we illustrate here \cite{chakrabarti2017PNAS}. It is noteworthy that despite the critical difference between CYT-19 and GroEL, with the former that disrupts both  the folded and misfolded states of ribozymes whereas the latter  does not interact with the folded proteins,  the mechanisms of their functions are in accord with the predictions of IAM.  Both GroEL and RNA chaperones function by driving the SPs and ribozymes out of equilibrium \cite{chakrabarti2017PNAS}. Remarkably, we showed by analyzing experimental data on ribozymes and MDH that the quantity that is optimized by GroEL and RNA chaperones is,
\begin{equation}
\Delta_{NE} = k_{F} P_N^{SS}
\label{optimum}
\end{equation}
where $k_{F}$ is the folding rate and $P_N^{SS}$ is the steady state yield (see Fig.\ref{Fig. 7Shaon17PNAS}). Thus, neither the folding rate nor the steady state yield is maximized but it is the product of the two that is optimized by the molecular chaperones. It follows from Fig.\ref{Fig. 7Shaon17PNAS} that, for a given SP and external conditions, which would fix $k_{F}$,  the steady state yield would have the largest value for
 the wild type GroEL than any other mutant.  That this is indeed the case is vividly illustrated in Fig. \ref{Tehver2a08JMB}. In the case of GroEL, the value of $P_N^{SS}$ (or $P_N(t)$ at any $t$) is critically dependent only on $k_{R^{\prime\prime} \rightarrow T}$, which has the largest value for the WT GroEL. The optimality condition given in Eq.\ref{optimum} is determined by the value of $k_{R^{\prime\prime} \rightarrow T}$, which in turn depends on the dynamics of allosteric transitions as well the presence of SP. Thus, the function of GroEL, and most likely CYT-19 and related RNA chaperones, cannot be understood without considering the details of the reaction cycle and how they are directly related to SP folding. The IAM theory, which accounts for all the complexities of the reaction cycle, explains the available experimental data quantitatively (see for example Fig. \ref{Tehver2a08JMB}) using a single parameter ($k_{R^{\prime\prime} \rightarrow T}$).

 \subsection{When it does SP folding occurs in the expanded GroEL cavity} 
 Does SP folding occur in the expanded cavity or in solution after ejection? This  question has unnecessarily plagued the discussion of GroEL-assisted folding, causing substantial confusion largely because of insistence by some that GroEL merely encapsulates the SP in the cavity until it reaches the native state with unit probability \cite{Horwich09FEBSLett}. Such an inference that GroEL is a passive Anfinsen cage has been made principally using experiments based on a single ring mutant (SR1) from which discharge of GroES and the SP occurs on a time scale of 300 minutes is erroneous. For starters, the life time of the encapsulated SP in the wild type (WT) cycling GroEL is about 2 seconds \cite{Ye13PNAS} that is four orders of magnitude shorter than the SR1 lifetime! Furthermore, neither the passive or active cage model can explain how the communication to discharge the ligands (ADP and the inorganic phosphate), GroES, and the folded SP takes place. 
 
 Does folding to the folded state occur within the cavity in the WT GroEL? We answer this question in the affirmative by using the following argument.  Assisted folding requires that the kinetic constraint, $k_F <  k_B$ be satisfied  where $k_B$ pseudo first order binding rate of SP to GroEL. In the opposite limit ($k_F > k_B$), which is relevant at low GroEL concentrations, folding is sufficiently rapid compared to diffusion controlled binding that the chaperonin machinery would not be needed. Thus, assuming that the kinetic constraint ($k_B \gg k_A$) is always satisfied for stringent substrates under non-permissive conditions then the SP upon ejection from the GroEL cavity, roughly every two seconds, rebinds (presumably to the same GroEL molecule) rapidly.  If the ejected SP is in the folded state then it would not be recognized by GroEL because the hydrophobic recognition motifs would no longer be solvent exposed. Thus, the fate of SP, which occurs by the KPM, is decided entirely within the cavity during the lifetime of its residence. Both folding and partitioning to the ensemble of misfolded states occur rapidly while the SP is encapsulated for a brief period in either chamber. 

We provide evidence to substantiate the physical arguments given above. The theory underlying IAM was used to obtain the parameters for the rates in the catalytic cycle and the intrinsic rates for assisted folding of RUBISCO. The time for RUBISCO molecules to reach the folded state by the fast track, $\tau_F = k_F^{-1} = 0.6$ s (Table 1 in \cite{Tehver08JMB}), which is less than the encapsulation time of about 2 seconds. This implies only the fast track RUBISCO molecules fold in the cavity because time for slow track Rubisco molecules $\tau_S (=k_S^{-1})$ to fold is about 333 minutes (Table 1 in \cite{Tehver08JMB}). The slow track molecule would rapidly rebind upon exiting the cavity, and the process is iterated multiple times till the majority of unfolded SPs reach the native state. One can use the same argument for reconstituting Citrate Synthase (CS)  using GroEL. The fits to the experimental data \cite{Sun2003JMB} in Fig. \ref{Tehver2a08JMB} yields $\tau_F = 0.6$ s whereas $\tau_S = 100$ minutes \cite{Tehver08JMB}, which again shows that KPM resulting in folded and misfolded states occurs while CS is encapsulated in the cavity.  Thus, we can conclude that when SP folding occurs it occurs in the expanded cavity. It is worth emphasizing that because the IAM theory takes into account the coupling between the events in the reaction cycle of GroEL and SP folding it naturally explains the allosteric communication needed for discharge of the SP, whether it is folded or not, and other ligands. However, only a very small fraction reaches the folded state in each cycle, and hence the need to perform the iterations as rapidly as possible. Remarkably, GroEL has evolved to do just that by functioning as a parallel processing machine in the symmetric complex when challenged with SP \cite{Yang13PNAS,Takei12JBC}.

\subsection{Symmetric Complex is the Functioning Unit of the GroEL-GroES machine} 
The IAM predicts that the yield of the folded SP increases with each iteration. It, therefore, follows that for highly efficacious function  it would be optimal if GroEL-GroES functions as a parallel processing machine with one SP in each chamber. This would necessarily involve formation of a symmetric complex GroEL$_{14}$-GroES$_{14}$, which was  shown as the functioning unit only recently \cite{Takei12JBC,Fei14PNAS,Ye13PNAS}. In particular, using a FRET-based system Ye and Lorimer \cite{Ye13PNAS} have established unequivocally that the response of the GroEL-GroES machinery is dramatically different  with and without the presence of SP. In order to unveil the differences they had to follow the fate of ADP and P$_i$ release in real time. These experiments showed that in the absence of the SP the rate determining step involves release of P$_i$ {\it before} ADP release from the {\it trans} ring of the dominant asymmetric complex (GroES bound to the {\it cis} ring).  In sharp contrast, when challenged with the SP, ADP is released {\it before} P$_i$. The symmetric particle, with GroES bound to both the rings (Fig.\ref{structure}b), is the predominant species in the presence of SP. In principle, the symmetric particle can simultaneously facilitate the folding of two SPs one in each chamber. Thus, it is likely the case that the functional form {\it in vivo} is the symmetric particle , which is activated when there is a job to do, namely, help SPs fold.

There was one other major finding in the Ye-Lorimer study \cite{Ye13PNAS}. They discovered that the ATP hydrolysis rate ($\sim$ 0.5$^{-1}$) is the same in the presence and absence of the SP. In the presence of SP, hydrolysis of ATP is rate limiting, which in the language used to describe motility of motors means that GroEL is ATP-gated. In other words, symmetry breaking (or inter ring communication) events that determine the ring from which GroEL disassociates depends on extent of ATP hydrolysis in each ring. Remarkably, the release of ADP from the {\it trans} ring is accelerated roughly 100 fold in the presence of SP. We note parenthetically that release of ADP from the nucleotide binding pocket of conventional kinesin is accelerated by nearly 1000 fold in the presence of microtubules \cite{Hackney88PNAS}, hinting at the possibility that there is a unified molecular basis for nucleotide chemistry in biological machines.  By greatly enhancing ADP release in the presence of SP, resetting to the initial SP accepting state occurs rapidly ($k_{R^{\prime\prime} \rightarrow T}$ is maximized in the WT GroEL), which allows GroEL to process as many SP molecules as fast as possible. Clearly, these findings are in complete accord with the IAM predictions and debunk the Anfinsen cage model \cite{Horwich09FEBSLett,BrinkerCell01}. 

\section{Conclusions}
In this perspective, we have shown that, despite profound differences, the functions of GroEL-GroES machine and RNA chaperones are quantitatively described by the theory underlying the Iterative Annealing Mechanism. We are unaware any experiment of assisted folding of the SPs or ribozymes that cannot be explained by the theory. We conclude with the following additional comments.
\begin{enumerate}
\item
It is sometimes stated that the mechanism of how GroEL functions is controversial because of the proposal that the cavity in the GroEL could act as an Anfinsen cage in which folding can be completed unhindered by aggregation. Such a conclusion was reached based mostly on experiments on the SR1 mutant (an asymmetric GroEL complex) from which GroES disassociates in 300 minutes.  Although experiments using SR1 (with ADP and P$_i$ locked into the equatorial domain) provide insights into effects of confinement on SP folding they are irrelevant in  understanding of WT GroEL function.   Finally, in the Anfinsen cage model there is no necessity for invoking allosteric transitions and how they are linked to assisted folding. In the SR1 mutant, ATP binding and hydrolysis occurs once, which means that the SP is trapped in a hydrophilic cavity for extremely long times, and hence lessons from the SR1 mutant neither inform us about the intact WT GroEL nor are they biologically relevant.  On the other hand, the stochastic WT GroEL comes alive when presented with SPs, undergoes a series of allosteric transitions by binding, hydrolyzing ATP, and releasing the products, which permits the SPs multiple chances to fold in the most optimal fashion (see Eq.\ref{optimum}). The quantitative success of the IAM should put to rest the inadequacy and the erroneous Anfinsen cage model \cite{Horwich09FEBSLett} for describing the function of the WT GroEL. For instance, the results in Fig. \ref{mtMDH_CS} cannot be understood within the Anfinsen cage model. 

\item
The machine-like non-equilibrium characteristics of chaperones are most evident by the demonstration that the steady state yield, $P_N^{SS} \ne P_{EQ}$ where the equilibrium yield of the folded state, $P_{EQ}$, is given by the Boltzmann distribution,
\begin{equation}
P_N^{EQ} = \frac{1}{(1 + e^{-\Delta G_{NM}/k_BT})},
\label{Boltzmann}
\end{equation}
where $\Delta G_{NM}$ is the free energy of the folded state with respect to the manifold of misfolded states.  The values of $P_N^{SS}$ for the two SPs and {\it Tetrahymena} ribozyme and its variants depend both the chaperone and ATP concentrations \cite{chakrabarti2017PNAS}, which itself is evidence of departure from equilibrium. In addition, the measured value of $\Delta G_{NM}$ for the WT ribozyme is $\sim$ 10 $k_B T$, which implies that $P_N^{EQ} \approx$ 0.99 according to Eq. \ref{Boltzmann}. However, the measured value in the presence of ATP is far less, which shows that $P_N^{SS} \ne P^{EQ}_N$. The finding that $P_N^{SS}$ values of RUBISCO and MDH are dependent on GroEL concentration also implies that in the presence of GroEL Eq. \ref{Boltzmann} is not valid. Taken together they imply that in the process of assisted folding both GroEL and CYT-19 drive the misfolded SPs and ribozymes out of equilibrium (see also \cite{Goloubinoff18NCB}). 

\item
GroEL and RNA chaperones burn copious amount of ATP because in each round only a small fraction ($\Phi \ll 1$) of misfolded molecules reach the native state.  Consider RUBISCO folding for which $\Phi=0.02$ \cite{Tehver08JMB}. The yield of the native state at $t=20$ min with the concentration of GroEL roughly equal to the initial unfolded RUBISCO (both at 50 nm) is about 0.7 (see Fig.\ref{Tehver2a08JMB}). The value of $P_N^{SS}\approx 0.8$ from which we obtain $n \approx$ 100 using Eq.\ref{yield}. In each catalytic cycle between $(3-4)$ ATP molecules are consumed, which implies that in order to fold roughly 88\% of RUBISCO in the steady state between $(300-400)$ ATP molecules are hydrolyzed. As pointed out elsewhere \cite{Todd96PNAS}, this is but a very small fraction of energy required to synthesize RUBISCO, a protein with 491 residues. Thus, the benefits of GroEL assisted folding far outweighs the cost of protein synthesis. However, from a thermodynamic perspective it can be argued that GroEL is less efficient than Myosin V, which consumes one ATP molecule (available energy is about $E_{ATP} \approx (20-25)$ $k_BT$) per step ($s \approx$ 36 nm) while walking on actin filament, resisting forces on the order of about $f_s \approx$ 2 pN. Thus, a rough estimate of Myosin V efficiency is $\frac{sf_s}{E_{ATP}}$ is very high.

\end{enumerate}

\section*{Acknowledgments} We are grateful to Bernard Brooks, Shaon Chakrabarti, Eda Koculi, George Stan, Riina Tehver, and Scott Ye for collaboration on various aspects of the works described here. DT acknowledges useful conversations with Alan Lambowitz on RNA chaperones. This work was supported in part by a grant from National Science Foundation (CHE 19-00093) and the Collie-Welch Chair (F-0019) administered through the Welch Foundation.


\begin{thebibliography}{10}

\bibitem{Lorimer89Nature}
Goloubinoff P, Gatenby AA, Lorimer GH
\newblock (1989) Gro{EL} heat-shock proteins promote assembly of foreign
  prokaryotic ribulose bisphosphate carboxylase oligomers in
  {E}scherichia-{C}oli.
\newblock Nature 337:44--47.

\bibitem{Sigler98AnnRevBiochem}
Sigler PB, Xu Z, Rye HS, Burston SG, Fenton WA, Horwich AL
\newblock (1998) Structure and function in groel-mediated protein folding.
\newblock Annu. Rev. Biochem. 67:581--608.

\bibitem{Thirumalai01ARBB}
Thirumalai D, Lorimer GH
\newblock (2001) Chaperonin-mediated protein folding.
\newblock Ann. Rev. Biophys. Biomol. Struct. 30:245--269.

\bibitem{Ellis87Nature}
Ellis J
\newblock (1987) Proteins as molecular chaperones.
\newblock Nature 328:378--379.

\bibitem{Lorimer01PlantPhysiol}
Lorimer GH
\newblock (2001) A personal account of chaperonin history.
\newblock Plant physiology 125:38--41.

\bibitem{Hemmingsen88Nature}
Hemmingsen SM, Woolford C, van~der Vies SM, Tilly K, Dennis DT,
  Georgopoulos CP, Hendrix RW,  Ellis RJ
\newblock (1988) Homologous plant and bacterial proteins chaperone oligomeric
  protein assembly.
\newblock Nature 333:330.

\bibitem{Mohr02Cell}
Mohr S, Stryker JM,  Lambowitz AM
\newblock (2002) A dead-box protein functions as an atp-dependent rna chaperone
  in group i intron splicing.
\newblock Cell 109:769--779.

\bibitem{Woodson10RNABiol}
Woodson SA
\newblock (2010) Taming free energy landscapes with rna chaperones.
\newblock RNA biology 7:677--686.

\bibitem{Pan1997JMB}
Pan J, Thirumalai D, Woodson SA
\newblock (1997) Folding of rna involves parallel pathways.
\newblock J. Mol. Biol. 273:7--13.

\bibitem{Thirumalai19ChemRev}
Thirumalai D, Hyeon C, Zhuravlev PI, Lorimer GH
\newblock (2019) Symmetry, rigidity, and allosteric signaling: From monomeric
  proteins to molecular machines.
\newblock Chem. Rev. 119:6788--6821.

\bibitem{Xu97Nature}
Xu Z, Horwich AL,  Sigler, PB
\newblock (1997) The crystal structure of the asymmetric
  {GroEL-GroES-(ADP)$_7$} chaperonin complex.
\newblock Nature 388:741.

\bibitem{Fei13PNAS}
Fei X, Yang D, LaRonade-LeBlanc N, Lorimer GH
\newblock (2013) Crystal structure of a groel-adp complex in the relaxed
  allosteric state at 2.7 \AA\ resolution.
\newblock Proc. Natl. Acad. Sci. U. S. A. 110:E2958--E2966.

\bibitem{Fei14PNAS}
Fei X, Ye X, LaRonade NA, Lorimer GH
\newblock (2014) Formation and structures of groel:groes2 chaperonin footballs,
  the protein-folding functioal form.
\newblock Proc. Natl. Acad. Sci. U. S. A. 111:12776--12780.

\bibitem{Todd96PNAS}
Todd MJ, Lorimer GH,  Thirumalai D
\newblock (1996) Chaperonin-facilitated protein folding: Optimization of rate
  and yield by an iterative annealing mechanism.
\newblock Proc. Natl. Acad. Sci. U. S. A. 93:4030--4035.

\bibitem{Gruber16ChemRev}
Gruber R, Horovitz A
\newblock (2016) Allosteric mechanisms in chaperonin machines.
\newblock Chem. Rev. 116:6588--6606.

\bibitem{Thirumalai18PhilTrans}
Thirumalai D, Hyeon C
\newblock (2018) {Signalling networks and dynamics of allosteric transitions in
  bacterial chaperonin GroEL: implications for iterative annealing of misfolded
  proteins}.
\newblock Phil. Trans. Royal Soc. B: Biol. Sci. 373:20170182.

\bibitem{Gruber18PhilTransRoySocB}
Gruber R, Horovitz A
\newblock (2018) Unpicking allosteric mechanisms of homo-oligomeric proteins by
  determining their successive ligand binding constants.
\newblock Phil. Trans. R. Soc. B 373:20170176.

\bibitem{Yang13PNAS}
Yang D, Ye X, Lorimer GH
\newblock (2013) {Symmetric GroEL:GroES2 complexes are the protein-folding
  functional form of the chaperonin nanomachine}.
\newblock Proc. Natl. Acad. Sci. U. S. A. 110:E4298--E4305.

\bibitem{Takei12JBC}
Takei Y, Iizuka R, Ueno T, Funatsu T
\newblock (2012) {Single-molecule observation of protein folding in symmetric
  GroEL-(GroES) 2 complexes}.
\newblock J. Biol. Chem. 287:41118--41125.

\bibitem{Viitanen92ProtSci}
Viitanen PV, Gatenby AA, Lorimer GH
\newblock (1992) Purified chaperonin 60 (groel) interacts with the nonnative
  states of a multitude of escherichia coli proteins.
\newblock Protein Science 1:363--369.

\bibitem{Betancourt99JMB}
Betancourt MR, Thirumalai D
\newblock (1999) Exploring the kinetic requirements for enhancement of protein
  folding rates in the {GroEL} cavity.
\newblock J. Mol. Biol. 287:627--644.

\bibitem{Baumketner03JMB}
Baumketner A, Jewett A, Shea J
\newblock (2003) Effects of confinement in chaperonin assisted protein folding:
  rate enhancement by decreasing the roughness of the folding energy landscape.
\newblock Journal of molecular biology 332:701--713.

\bibitem{Cheung06JMB}
Cheung MS, Thirumalai D
\newblock (2006) Nanopore--protein interactions dramatically alter stability
  and yield of the native state in restricted spaces.
\newblock J. Mol. Biol. 357:632--643.

\bibitem{Hoffmann10PNAS}
Hofmann H, Hillger F, Pfeil SH, Hoffmann A, Streich D, Haenni D,
  Nettels D, Lipman EA,  Schuler B
\newblock (2010) Single-molecule spectroscopy of protein folding in a
  chaperonin cage.
\newblock Proceedings of the National Academy of Sciences 107:11793--11798.

\bibitem{Fedorov97JMB}
Fedorov AN, Baldwin TO
\newblock (1997) {GroE modulates kinetic partitioning of folding intermediates
  between alternative states to maximize the yield of biologically active
  protein}.
\newblock J. Mol. Biol. 268:712--723.

\bibitem{Treiberscience98}
Treiber DK, Rook M, Zarrinkar PR, Williamson, JR
\newblock (1998) {Kinetic Intermediates Trapped by Native Interactions in RNA
  Folding}.
\newblock Science 279:1943--1946.

\bibitem{Treiber99COSB}
Treiber DK, Williamson JR
\newblock (1999) {Exposing the kinetic traps in RNA folding}.
\newblock Curr. Opin. Struct. Biol. 9:339--345.

\bibitem{Thirumalai96ACR}
Thirumalai D, Woodson SA
\newblock (1996) {Kinetics of Folding of Proteins and RNA}.
\newblock Acc. Chem. Res. 29:433--439.

\bibitem{RussellJMB01}
Russell R, Herschlag D
\newblock (2001) Probing the folding landscape of the tetrahymena ribozyme:
  Commitment to form the native conformation is late in the folding pathway.
\newblock J. Mol. Biol. 308:839--851.

\bibitem{Woodson05COSB}
Woodson SA
\newblock (2005) Structure and assembly of group i introns.
\newblock Curr. Opin. Struct. Biol. 15:324--330.

\bibitem{RussellJMB99}
Russell R, Herschlag D
\newblock (1999) New pathways in folding of the \emph{Tetrahymena} group {I}
  {RNA} enzyme.
\newblock J. Mol. Biol. 291:1155--1167.

\bibitem{Thirum05Biochem}
Thirumalai D, Hyeon C
\newblock (2005) {RNA and Protein folding: Common Themes and Variations}.
\newblock Biochemistry 44:4957--4970.

\bibitem{Hyeon2014IJC}
Hyeon C, Denesyuk, NA, Thirumalai D
\newblock (2014) {Development and Applications of Coarse-Grained Models for
  RNA}.
\newblock Israel J. Chem. 54:1358--1373.

\bibitem{Solomatin10Nature}
Solomatin SV, Greenfeld M, Chu S, Herschlag D
\newblock (2010) Multiple native states reveal persistent ruggedness of an
  {RNA} folding landscape.
\newblock Nature 463:681--684.

\bibitem{ThirumARPC01}
Thirumalai D, Lee N, Woodson SA, Klimov DK
\newblock (2001) {Early Events in RNA Folding}.
\newblock Annu. Rev. Phys. Chem. 52:751--762.

\bibitem{Zhuang00Science}
Zhuang X, Bartley L, Babcock A, Russell R, Ha T, Hershlag D, Chu S
\newblock (2000) A single-molecule study of {RNA} catalysis and folding.
\newblock Science 288:2048--2051.

\bibitem{Pan00JMB}
Pan J, Deras ML, Woodson SA
\newblock (2000) {Fast Folding of a Ribozyme by Stabilization Core
  Interactions: Evidence for Multiple Folding Pathways in RNA}.
\newblock J. Mol. Biol. 296:133--144.

\bibitem{Lorsch2002Cell}
Lorsch JR
\newblock (2002) Rna chaperones exist and dead box proteins get a life.
\newblock Cell 109:797--800.

\bibitem{Tijerina2006PNAS}
Tijerina P, Bhaskaran H, Russell R
\newblock (2006) Nonspecific binding to structured {RNA} and preferential
  unwinding of an exposed helix by the {CYT-19} protein, a {DEAD-box} {RNA}
  chaperone.
\newblock Proc. Natl. Acad. Sci. U. S. A. 103:16698--16703.

\bibitem{Bhaskaran07Nature}
Bhaskaran H, Russell R
\newblock (2007) Kinetic redistribution of native and misfolded rnas by
  dead-box chaperone.
\newblock Nature 449:1014--1018.

\bibitem{Grohman2007Biochemistry}
Grohman JK, Del~Campo M, Bhaskaran H, Tijerina P, Lambowitz AM,
  Russell R
\newblock (2007) {Probing the mechanisms of DEAD-box proteins as general RNA
  chaperones: the C-terminal domain of CYT-19 mediates general recognition of
  RNA}.
\newblock Biochemistry 46:3013--3022.

\bibitem{Mallam2011PNAS}
Mallam AL, Jarmoskaite I, Tijerina P, Del~Campo M, Seifert S, Guo L,
  Russell R,  Lambowitz AM.
\newblock (2011) Solution structures of dead-box rna chaperones reveal
  conformational changes and nucleic acid tethering by a basic tail.
\newblock Proc. Natl. Acad. Sci. U. S. A. 108:12254--12259.

\bibitem{Russell2013RNAbiology}
Russell R, Jarmoskaite I, Lambowitz AM
\newblock (2013) {Toward a molecular understanding of RNA remodeling by
  DEAD-box proteins}.
\newblock RNA Biology 10:44--55.

\bibitem{lorimer1996FASEBJ}
Lorimer GH
\newblock (1996) {A quantitative assessment of the role of the chaperonin
  proteins in protein folding in vivo}.
\newblock FASEB J. 10:5.

\bibitem{Stan05ProtSci}
Stan G, Brooks BR, Lorimer GH,  Thirumalai D
\newblock (2005) Identifying natural substrates for chaperonins using a
  sequence-based approach.
\newblock Prot. Sci. 14:193--201.

\bibitem{Chaudhuri05CellStress}
Chaudhuri TK, Gupta P
\newblock (2005) Factors governing the substrate recognition by groel
  chaperone: a sequence correlation approach.
\newblock Cell stress \& chaperones 10:24.

\bibitem{Stan06PNAS}
Stan G, Brooks BR, Lorimer GH,  Thirumalai D
\newblock (2006) Residues in substrate proteins that interact with groel in the
  capture process are buried in the native state.
\newblock Proc. Natl. Acad. Sci. U. S. A. 103:4433--4438.

\bibitem{Noivirt-Brik07Bioinformatics}
Noivirt-Brik O, Unger R,  Horovitz A
\newblock (2007) {Low folding propensity and high translation efficiency
  distinguish in vivo substrates of GroEL from other Escherichia coli
  proteins}.
\newblock Bioinformatics 23:3276--3279.

\bibitem{Endo07BiosciBiotechnolBiochem}
Endo A, Kurusu Y
\newblock (2007) {Identification of in vivo substrates of the chaperonin GroEL
  from Bacillus subtilis}.
\newblock Biosci. Biotech. Biochem. 71:1073--1077.

\bibitem{Guo95Biopolymers}
Guo Z, Thirumalai D
\newblock (1995) {Kinetics of Protein Folding: Nucleation Mechanism, Time
  Scales, and Pathways}.
\newblock Biopolymers 36:83--102.

\bibitem{Thirumalai97TCA}
Thirumalai D, Klimov DK, Woodson SA
\newblock (1997) Kinetic partitioning mechanism as a unifying theme in the
  folding of biomolecules.
\newblock Theor. Chem. Acc. 96:14--22.

\bibitem{Thirum95JPI}
Thirumalai D
\newblock (1995) {From Minimal Models to Real Proteins: Time Scales for Protein
  Folding Kinetics}.
\newblock J. Phys. I (Fr.) 5:1457--1467.

\bibitem{Ziv09PCCP}
Ziv G, Thirumalai D, Haran G
\newblock (2009) Collapse transition in proteins.
\newblock Phys. Chem. Chem. Phys. 11:83--93.

\bibitem{Schmidt94JBC}
Schmidt M, Buchner J, Todd MJ, Lorimer GH, Viitanen PV
\newblock (1994) On the role of groes in the chaperonin-assisted folding
  reaction. three case studies.
\newblock J. Biol. Chem. 269:10304--10311.

\bibitem{Mattingly95JBC}
Mattingly JR, Iriarte A, Martinez-Carrion M
\newblock (1995) Homologous proteins with different affinities for groel the
  refolding of the aspartate aminotransferase isozymes at varying temperatures.
\newblock J. Biol. Chem. 270:1138--1148.

\bibitem{Kiefhaber95PNAS}
Kiefhaber T
\newblock (1995) Kinetic traps in lysozyme folding.
\newblock Proc. Natl. Acad. Sci. U. S. A. 92:9029--9033.

\bibitem{Matagne1998JMB}
Matagne A, Chung EW, Ball LJ, Radford SE, Robinson CV, Dobson
  CM
\newblock (1998) {The origin of the $\alpha$-domain intermediate in the folding
  of hen lysozyme}.
\newblock J. Mol. Biol. 277:997--1005.

\bibitem{Li08PNAS}
Peng Q, Li H
\newblock (2008) Atomic force microscopy reveals parallel mechanical unfolding
  pathways of t4 lysozyme: evidence for a kinetic partitioning mechanism.
\newblock Proc. Natl. Acad. Sci. U. S. A. 105:1885--1890.

\bibitem{Stigler11Science}
Stigler J, Ziegler F, Gieseke A, Gebhardt J, Rief M
\newblock (2011) {The Complex Folding Network of Single Calmodulin Molecules}.
\newblock Science 334:512--516.

\bibitem{Dima04JPCB}
Dima RI, Thirumalai D
\newblock (2004) Asymmetry in the shapes of folded and denatured states of
  proteins.
\newblock J. Phys. Chem. B 108:6564--6570.

\bibitem{Lorimer18PhilTrans}
Lorimer GH, Horovitz A, McLeish T
\newblock (2018) Allostery and molecular machines.
\newblock Philos Trans R Soc Lond B Biol Sci. 373:20170173.

\bibitem{Sameshima10BiochemJ}
Sameshima T, Iizuka R, Ueno T, Funatsu T
\newblock (2010) {Denatured proteins facilitate the formation of the
  football-shaped GroEL--(GroES) 2 complex}.
\newblock Biochem. J. 427:247--254.

\bibitem{Ye13PNAS}
Ye X, Lorimer GH
\newblock (2013) Substrate protein switches groe chaperonins from asymmetric to
  symmetric cycling by catalyzing nucleotide exchange.
\newblock Proc. Natl. Acad. Sci. U. S. A. 110:E4289--E4297.

\bibitem{Horovitz01JSB}
Horovitz A, Fridmann Y, Kafri G, Yifrach O
\newblock (2001) Review:allostery in chaperonins.
\newblock J. Struct. Biol. 135:104--114.

\bibitem{Yifrach95Biochem}
Yifrach O, Horovitz A
\newblock (1995) Nested cooperativity in the {ARP}ase activity of the
  oligomeric chaperonin {GroEL}.
\newblock Biochemistry 34:5303--5308.

\bibitem{Yifrach00PNAS}
Yifrach O, Horovitz A
\newblock (2000) Coupling between protein folding and allostery in the groe
  chaperonin system.
\newblock Proc. Natl. Acad. Sci. U. S. A. 97:1521--1524.

\bibitem{Tehver09JMB}
Tehver R, Chen J, Thirumalai D
\newblock (2009) Allostery wiring diagrams in the transitions that drive the
  groel reaction cycle.
\newblock J. Mol. Biol. 387:390--406.

\bibitem{ThirumalaiBookDoniach}
Thirumalai D
\newblock (1994) {\em {Statistical Mechanics, Protein Structure, and Protein -
  Substrate Interactions}}, ed.{} Doniach, S.
\newblock (Plenum, New York), pp. 115 -- 134.

\bibitem{Orland97JP}
Orland H, Thirumalai D.
\newblock (1997) A kinetic model for chaperonin assisted folding of proteins.
\newblock J. Phys. I France 7:553--560.

\bibitem{Hyeon06PNAS}
Hyeon C, Lorimer GH,  Thirumalai D
\newblock (2006) {Dynamics of allosteric transition in GroEL}.
\newblock Proc. Natl. Acad. Sci. U. S. A. 103:18939--18944.

\bibitem{Corsepius13PNAS}
Corsepius NC, Lorimer GH
\newblock (2013) Measuring how much work the chaperone groel can do.
\newblock Proc. Natl. Acad. Sci. U. S. A. 110:E2451--E2459.

\bibitem{Anfinsen73Science}
Anfinsen CB
\newblock (1973) Principles that govern the folding of protein chain.
\newblock Science 181:223--230.

\bibitem{Kirkpatrick83Science}
Kirkpatrick S, Gelatt D, Vecchi MP
\newblock (1983) Optimization by simulated annealing.
\newblock Science 220:671--680.

\bibitem{Tehver08JMB}
Tehver R, Thirumalai D
\newblock (2008) {Kinetic Model for the Coupling between Allosteric Transitions
  in GroEL and Substrate Protein Folding and Aggregation}.
\newblock J. Mol. Biol. 377:1279--1295.

\bibitem{chakrabarti2017PNAS}
Chakrabarti S, Hyeon C, Ye X, Lorimer G, Thirumalai D
\newblock (2017) {Molecular Chaperones Maximize the Native State Yield on
  Biological Times by Driving Substrates out of Equilibrium}.
\newblock Proc. Natl. Acad. Sci. U. S. A. 114:E10919--E10927.

\bibitem{Sun2003JMB}
Sun Z, Scott DJ, Lund PA
\newblock (2003) Isolation and characterisation of mutants of groel that are
  fully functional as single rings.
\newblock J. Mol. Biol. 332:715--728.

\bibitem{Hyeon2013JCP}
Hyeon C, Thirumalai D
\newblock (2013) Generalized iterative annealing model for the action of rna
  chaperones.
\newblock J. Chem. Phys. 139:121924.

\bibitem{Horwich09FEBSLett}
Horwich AL, Apetri AC, Fenton WA
\newblock (2009) The groel/groes cis cavity as a passive anti-aggregation
  device.
\newblock FEBS letters 583:2654--2662.


\bibitem{Goloubinoff18NCB}
Goloubinoff P, Sassi AS, Fauvet B, Barducci A and De los Rios P
\newblock (2018) Chaperones convert the energy from ATP into the nonequilibrium
   stabilization of native proteins.
\newblock Nat. Chem. Biol. 14:388--394.

\bibitem{Hackney88PNAS}
Hackney DD
\newblock (1988) Kinesin atpase: rate-limiting adp release.
\newblock Proceedings of the National Academy of Sciences 85:6314--6318.

\bibitem{BrinkerCell01}
Brinker A, Pfeifer G, Kerner MJ, Naylor DJ, Hartl FU, 
  {Hayer-Hartl} M
  \newblock (2001) {Dual Function of Protein Confinement in Chaperonin-Assisted
  Protein Folding}.
\newblock Cell 107:223--233.

\end{thebibliography}

\clearpage

\end{document}